\documentclass[aps,prl,groupedaddress,twocolumn,superscriptaddress,floatfix,10pt]{revtex4-2}

\usepackage{xcolor}
\usepackage{color,soul}
\usepackage{graphicx}
\usepackage{mathtools}
\usepackage{siunitx}
\usepackage{float}
\usepackage{amssymb}
\begin{document}


\title{Decoupling of the many-body effects from the electron mass in GaAs by means of reduced dimensionality}

\author{P.~M.~T. Vianez}
\affiliation{Department of Physics, Cavendish Laboratory, University of Cambridge, Cambridge, CB3 0HE, UK}
\author{Y.~Jin}
\affiliation{Department of Physics, Cavendish Laboratory, University of Cambridge, Cambridge, CB3 0HE, UK}
\author{W.~K. Tan}
\affiliation{Department of Physics, Cavendish Laboratory, University of Cambridge, Cambridge, CB3 0HE, UK}
\author{Q.~Liu}
\affiliation{Department of Physics, Cavendish Laboratory, University of Cambridge, Cambridge, CB3 0HE, UK}
\author{J.~P. Griffiths}
\affiliation{Department of Physics, Cavendish Laboratory, University of Cambridge, Cambridge, CB3 0HE, UK}
\author{I. Farrer}
\affiliation{Department of Electronic \& Electrical Engineering, University of Sheffield, Mappin Street, Sheffield, S1 3JD, UK}
\author{D.~A. Ritchie}
\affiliation{Department of Physics, Cavendish Laboratory, University of Cambridge, Cambridge, CB3 0HE, UK}
\author{O. Tsyplyatyev}
\affiliation{Institut f\"ur Theoretische Physik, Universit\"at Frankfurt, Max-von-Laue Stra{\ss}e 1, 60438 Frankfurt, Germany}
\author{C.~J.~B. Ford}
\affiliation{Department of Physics, Cavendish Laboratory, University of Cambridge, Cambridge, CB3 0HE, UK}


\begin{abstract}
Determining the (bare) electron mass $m_0$ in crystals is often hindered by many-body effects since Fermi-liquid physics renormalises the band mass, making the observed effective mass $m^*$ depend on density. Here, we use a one-dimensional (1D) geometry to amplify the effect of interactions, forcing the electrons to form a nonlinear Luttinger liquid with separate holon and spinon bands, therefore separating the interaction effects from $m_0$. Measuring the spectral function of gated quantum wires formed in GaAs by means of magnetotunnelling spectroscopy and interpreting them using the 1D Fermi-Hubbard model, we obtain $m_0=(0.0525\pm0.0015)m_\textrm{e}$ in this material, where $m_\textrm{e}$ is the free-electron mass. By varying the density in the wires, we change the interaction parameter $r_\textrm{s}$ in the range from $\sim$1--4 and show that $m_0$ remains constant. The determined value of $m_0$ is $\sim 22$\% lighter than observed in GaAs in geometries of higher dimensionality $D$ ($D>1$), consistent with the quasi-particle picture of a Fermi liquid that makes electrons heavier in the presence of interactions.
\end{abstract}


\maketitle

\section{}

\section*{\textbf{I. Introduction}}

Since its creation in the 1920s \cite{goldschmidt_crystal_1929}, gallium arsenide (GaAs) has become one of the first materials of choice for studying a number of problems in fundamental physics, from the well-known quantum Hall effects \cite{klitzing_new_1980,tsui_two-dimensional_1982,laughlin_anomalous_1983}, to spin-orbit coupling \cite{winkler_2003}, and Wigner crystallisation \cite{monarkha_two-dimensional_2012}. Simultaneously, it has also been used, together with other compounds of the III-V family, in manufacturing a range of electronic devices, such as laser diodes, integrated circuits, and solar cells, with its versatility as a direct band-gap material being still reflected to date, as the most widely used semiconductor after silicon \cite{manes_gallium_2005}. One of the basic parameters of any material is the band mass of its electrons $m_0$, the value of which for GaAs is often quoted as the effective bulk (three-dimensional, 3D) mass of $m_{\textrm{3D}}^{*}=0.067m_{\textrm{e}}$
measured at low densities. Indeed, it is well-established that in a crystal the effective mass can often differ from its free-space counterpart by up to several orders of magnitude, something which is understood as a direct result of the electron wave function interfering with the ionic lattice. Additional degrees of freedom such as phonons, spin waves, and plasmons, as well as impurity scattering and spin-orbit interactions, have also been known to affect the effective mass of carriers, including bulk GaAs \cite{raymond_electron_1979}. At a deeper level, however, one may wonder how strong the effect of the unavoidable electron-electron (e-e) interactions may be on their mass, given that, according to Fermi-liquid theory \cite{Landau57}, this cannot be separated from the band-structure effect on the bare mass of one electron.

A way of controlling the effect of e-e interactions on the carrier mass is by altering the coordination number of the electrons, via lowering the dimensionality $D$ of the system. In 1970, after Esaki
and Tsu's \cite{esaki_superlattice_1970} breakthrough with the invention
of semiconductor quantum wells, two-dimensional electron systems became available, which have since been perfected to extremely high qualities. 
The study of the electron mass as a function of carrier density
in GaAs/AlGaAs two-dimensional (2D) heterostructures has, however,
resulted in conflicting results, with values both above and below the
band mass being reported across a range of techniques and for varying
carrier densities \cite{coleridge_effective_1996,hayne_exchange_1992,hatke_evidence_2013,tan_measurements_2005}.
Going further to a one-dimensional (1D) geometry changes the effect of interactions drastically, with reduced masses having already been found in gold atomic chains \cite{nilius_development_2002}. In semiconductor systems, however, the most drastic departure from the single-electron picture has been what came to be known as spin-charge separation \cite{Auslaender05,jompol_probing_2009} predicted by the Luttinger-liquid theory \cite{Tomonaga50,Luttinger63}. As a result, we have shown that the Fermi sea
of electrons described by only one mass (making the band-structure and the many-body effects fundamentally indistinguishable in $D>1$) separates into two
bands for excitations of spin (\emph{i.e.}, spinons) and charge (\emph{i.e.},
holons), which can be described by two incommensurate masses $m_{\textrm{s}}$ and
$m_{\textrm{c}}$, respectively \cite{vianez_observing_2021}. This offers a method for decoupling
the effect of interactions from the measured band mass, whereby $m_{0}$ can be determined by increasing
the density and observing the point where the two masses converge
in the non-interacting limit.

In the present work, we demonstrate the experimental feasibility of this method
in GaAs by probing the dispersion
of the system via a tunnelling spectroscopy technique.  The paper is organised as follows. In Sec.\, II we introduce the experimental devices and spectroscopy technique used to probe the dispersion of a 1D wire array. Since the same devices also allow for a 2D Fermi-liquid system to be probed, we discuss in Sec.\ III how our technique already results in mass values closely matching others obtained independently. In Sec.\ IV we then introduce the Hubbard model, used in analysing the 1D experimental data, and extract $m_0$ in Sec.\ V. Section\,VI discusses the results, comparing them with previous works at higher dimensionality before the conclusion in Sec.\ VII. Appendix A contains details on sample preparation and the experimental setup. Appendix B discusses the interaction parameter $r_\textrm{s}$ and how it can be enhanced by mapping at the bottom of the second 1D subband.

\section*{II. Experimental Setup}

The system we have investigated is composed of an array of gated 1D wires, separated from a nearby two-dimensional electron gas (2DEG) by a superlattice barrier. The surface structure of all devices was fabricated on a 200\,$\mu$m-wide etched Hall bar, with current flowing along the high-mobility axis ($\langle110\rangle$ direction) of a MBE-grown GaAs/Al$_{0.33}$Ga$_{0.67}$As double-well heterostructure. Electrical contact to both wells was established using AuGeNi ohmic contacts. All gates were patterned using electron-beam lithography, and consisted of split (SG), mid-line (MG), bar (BG), and cut-off (CG) gates (used to set up independent electrical contact to each well and, therefore, the tunnelling conditions), together with an array of air-bridge-connected \cite{jin_microscopic_2021} wire gates (WGs) (used to define the quantum wires in the top well only), see Fig.\ \ref{fig_device}a. For full details on sample preparation and measurement, see Appendix A. 

In order to extract the bare electron mass $m_0$, decoupled from e-e interaction effects, we perform a low-noise, low-temperature spectroscopy measurement of the tunnelling current between each layer, given by
\begin{align}
    I\propto\int {\textrm{d}\textbf{k}}\textrm{d}E\left[f_T(E-E_{\rm{F1D}}-eV_\textrm{DC})-f_T(E-E_{\textrm{F2D}})\right]\nonumber\\ 
    \times A_{1}({\bf k},E)A_{2}({\textbf{k}}+ed({\textbf{n}}\times{\textbf{B}})/\hbar,E-eV_{\textrm{DC}}),
\end{align}
Here, $e$ is the electronic charge, $f_\textrm{T}(E)$ the Fermi-Dirac distribution, $d$ the centre-to-centre wavefunction separation, $\textbf{n}$ the unit normal to the 2D plane, \textbf{B}=-B\textbf{$\hat{y}$} the magnetic-field vector, \textbf{$\hat{y}$} the unit vector in the $y$-direction, and $A_1(\textbf{k},E)$ and $A_2(\textbf{k},E)$ the spectral functions of the 1D and 2D systems, respectively, with Fermi energies $E_\textrm{F1D}$ and $E_\textrm{F2D}$. As can be seen, the tunnelling current is then proportional to the overlap integral of the two spectral functions. 

\begin{figure}[h]
    \centering
	\includegraphics[width=0.85\columnwidth]{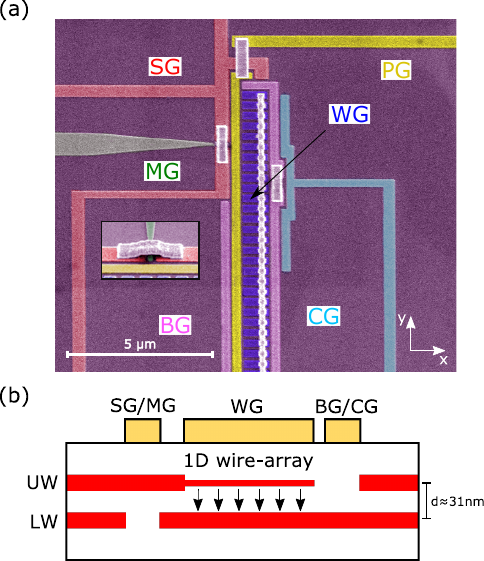}
    \caption{\textbf{A vertical tunnelling spectrometer.} (a) Scanning electron microscopy (SEM) micrograph of a 1\,$\mu$m long device. A number of electron-beam-defined gates are used in setting up the tunnelling conditions, see Appendix A for details. Inset: air-bridge interconnections, suspended $\sim100$\,nm over the surface \cite{jin_microscopic_2021}. (b) Schematic of a device operating in tunnelling mode. Here, tunnelling occurs between a 1D wire array defined in the upper well (UW) and a 2D spectrometer in the lower well (LW).} 
\label{fig_device}
\end{figure}

In our devices, while the bottom 2DEG always remains 2D in nature, the top 2DEG has confined (1D) regions, in between the wire-gates (together with a small 2D `parasitic' injection region, coloured yellow). This means that we can use the bottom layer as a well-understood spectrometer in order to probe the 1D dynamics, along the wire, taking place in the layer above, see Fig.\ \ref{fig_device}b for a shematic representation. An offset $eV_{\textrm{DC}}$ between the Fermi energies of the two systems is obtained by applying a DC bias $V_{\textrm{DC}}$ between the layers. Similarly, a shift in momentum can also be achieved via a magnetic field of strength $B$ parallel to the 2DEG layers, with the Lorentz force then adding $\Delta k=edB$ to the momentum of the tunnelling electrons, where $d$ is the centre-to-centre wavefunction separation. The dispersion of each system is then mapped by measuring the differential tunnelling conductance $G=\textrm{d}I/\textrm{d}V$ between the systems, as both energy and momentum are varied. 

\begin{figure*}[t]
    \centering
	\includegraphics[width=\textwidth]{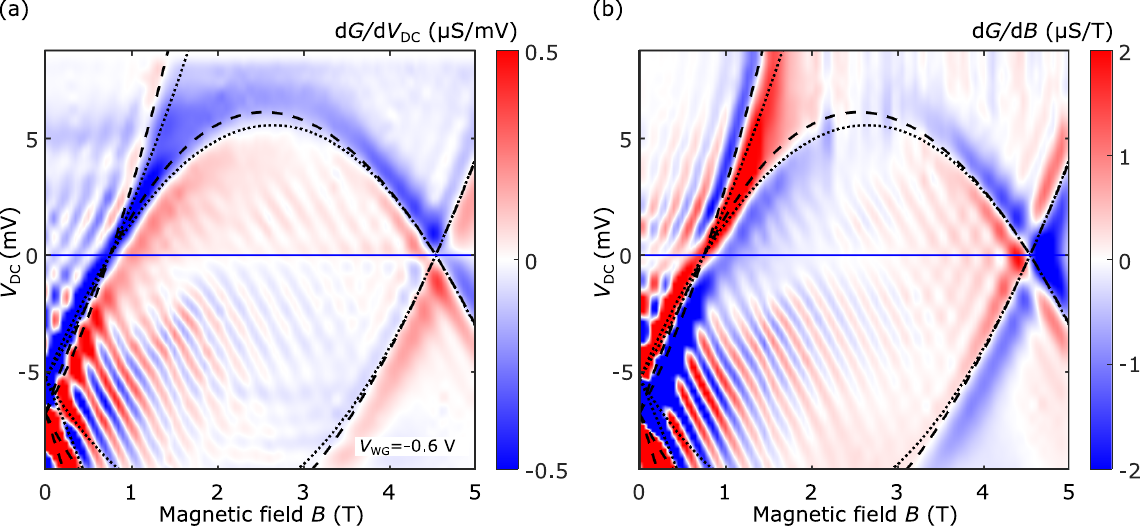}
    \caption{\textbf{Extraction of $m_\textrm{2D}^*$.} Tunnelling conductance differentials (a) $\textrm{d}G/\textrm{d}V_\textrm{DC}$ and (b) $\textrm{d}G/\textrm{d}B$ as a function of interlayer bias $V_\textrm{DC}$ and in-plane magnetic field $B$, for a device where the 1D wires were pinched off, so that only the 2D `parasitic' region is measured. Dashed and dotted black curves mark the dispersion of each 2DEG, respectively corrected and not corrected for capacitance, as given by Eqs.\ \ref{eq_parab} and \ref{eq_capacitance}. The extracted 2D mass value was $m_\textrm{2D}^{*}=(0.062\pm0.002)m_\textrm{e}$, in very good agreement with independent work already reported in the literature, see text.}
    \label{2D_2D_map}
\end{figure*}

\section*{III. Measuring $m_\textrm{2D}^*$}

We can set the wire-gate voltage $V_\textrm{WG}$ such that the wires pinch off and are unable to conduct. Under these conditions, electrons can only tunnel from the `parasitic' injection area running alongside the WG array. This is wide enough (0.45--0.6\,$\mu$m) for the electron gas to remain unconfined and, therefore, 2D in nature, see Fig.\ \ref{2D_2D_map}. Note that the electron densities in this region for both the upper (UW) and lower wells (LW) (see Fig.\ \ref{sfig1}c in Appendix B) are high enough for these systems to be treated as Fermi liquids, with the effective mass $m_\textrm{2D}^*$ renormalised by interactions.

The curves drawn in Fig.\ \ref{2D_2D_map} were obtained assuming single-electron tunnelling processes between the wells, and they mark the positions of resonant peaks arising from the maximal overlap of the offset spectral functions $\varepsilon(k)$ for both wells. These were obtained assuming a parabolic functional form
\begin{equation}
    \varepsilon_\textrm{2D}(k)=\frac{\hbar^2}{2m_\textrm{2D}^*}[k^2-(k_\textrm{F}^{\textrm{2D}})^2]
    \label{eq_parab}
\end{equation}
as well as conservation of energy and momentum during the tunnelling process
\begin{equation}
    \varepsilon_\textrm{UW}(k-\Delta k)=\varepsilon_\textrm{LW}(k)-eV_\textrm{DC},
    \label{eq_conservation}
\end{equation}
where $\Delta k=edB$.

From the MBE growth specifications we have simulated the expected band structure of the semiconductor wafer material used, see \cite{SM} for full details. The finite capacitance of the device also leads to an increase/decrease in the 2D electron density on each side of the barrier, $\pm\delta n_\textrm{2D}$, which in turn results in slightly asymmetric parabolae as $k_\textrm{F}^\textrm{2D}$ changes with interlayer voltage $V_\textrm{DC}$ ($\pm\delta k_\textrm{F}^\textrm{2D}=\pm\pi\delta n_\textrm{2D}/k_\textrm{F}^\textrm{2D}$). These can be modelled as
\begin{equation}
    k_\textrm{F}^{\textrm{2D'}}=k_\textrm{F}^\textrm{2D}\pm\delta k_\textrm{F}^\textrm{2D}=k_\textrm{F}^\textrm{2D}\pm\frac{\pi V_\textrm{DC}C}{ek_\textrm{F}^\textrm{2D}A},
    \label{eq_capacitance}
\end{equation}
where $C/A$ is the capacitance per unit area. 

The best match to the 2D-2D tunnelling signal coming from the `parasitic' injection region (shown in Fig.\ \ref{2D_2D_map}) was obtained with $d=31\,$nm, and capacitances $C_\textrm{UW}^\textrm{2D}=0.0047$\,Fm$^{-2}$ and $C_\textrm{LW}^\textrm{2D}=0.0033$\,Fm$^{-2}$ for the two wells. Note that the slight difference in capacitance between each well can be attributed to extra coupling arising from the different distances to the surface gates, with the ratio of depths $\sim 111/79=1.4$ and $C_\textrm{UW}^\textrm{2D}/C_\textrm{LW}^\textrm{2D}\sim 1.42$.
COMSOL \cite{COMSOL} simulations of our devices (which are not self-consistent) predict $C/A\approx 0.005$\,Fm$^{-2}$, very close to the values obtained from the fitting.

Having now accounted for well separation and capacitance effects, we obtain $m_\textrm{2D}^*=0.93m^*_\textrm{3D}=(0.062\pm0.002)m_\textrm{e}$, where $m^*_\textrm{3D}=0.067m_\textrm{e}$ is the electron mass in bulk GaAs in the low-density limit. This result is in very good agreement with independent work carried out in systems with similar densities and mobilities to ours, derived from both Shubnikov--de-Haas oscillations \cite{hayne_exchange_1992,coleridge_effective_1996,tan_measurements_2005}, microwave-induced resistance oscillations \cite{hatke_evidence_2013}, quantum Monte-Carlo calculations \cite{kwon_quantum_1994}, and cyclotron-resonance measurements \cite{kukushkin_fermi_2015}. 

\section*{IV. The Hubbard model}

\begin{figure*}[t]
    \centering
	\includegraphics[width=\textwidth]{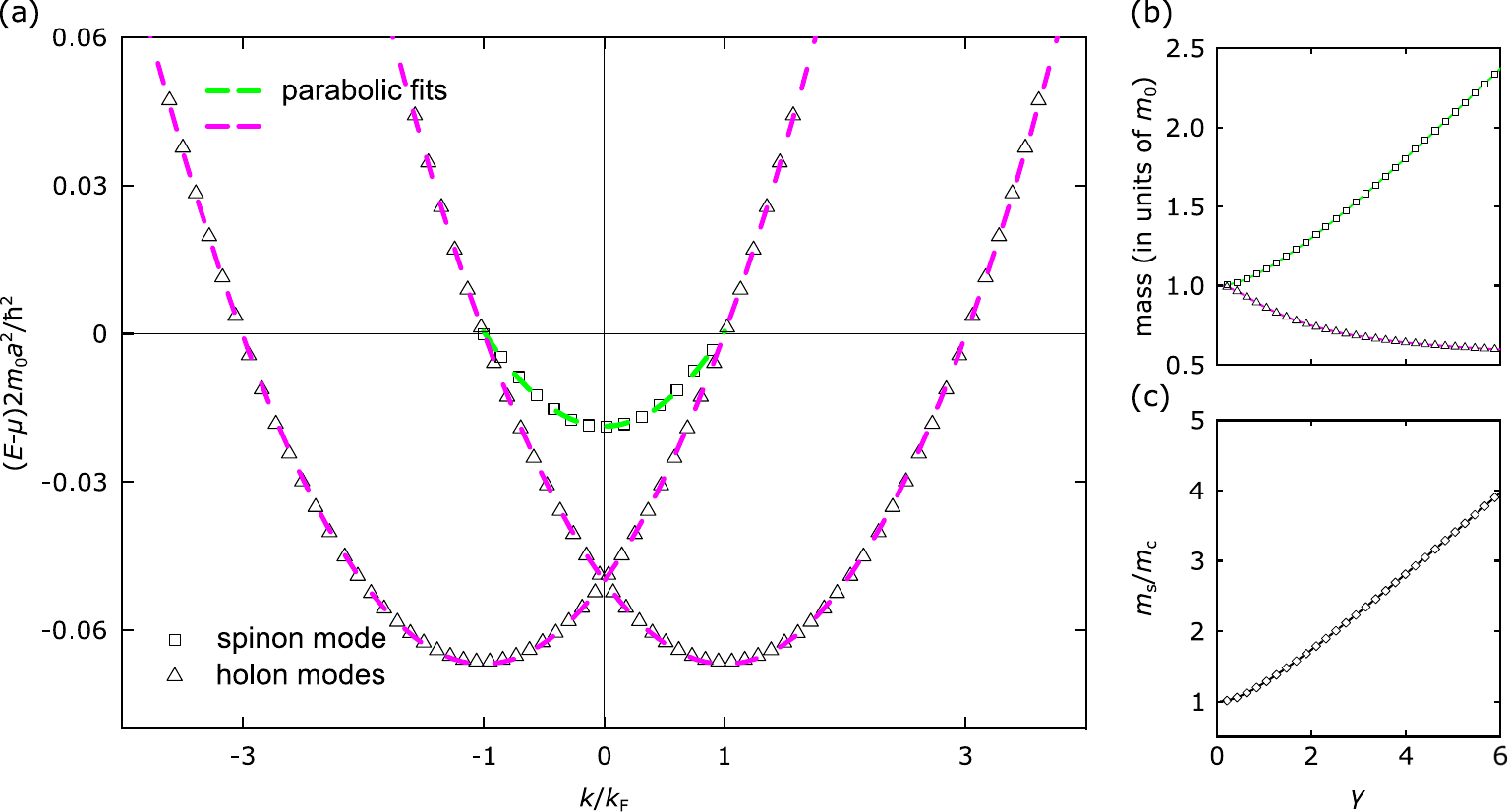}
    \caption{\textbf{1D many-body modes.} (a) Theoretically obtained spinon ($\Box$) and holon ($\triangle$) spectra as given by the Lieb-Wu equations for unpolarised electrons (i.e., M=N/2) for an interaction strength $\gamma=2$. The dashed curves correspond to parabolic fits, showing that each mode closely matches a parabola. From here we extract the respective spin $m_\textrm{s}$ and charge $m_\textrm{c}$ mass. Note that the charge mode crossing the line of the chemical potential at the $-3k_\textrm{F}$ and $+k_\textrm{F}$ points (or equivalently, at $-k_\textrm{F}$ and $+3k_\textrm{F}$) corresponds to the density of its constituent particles that is twice that for the spin mode crossing at the $\pm k_\textrm{F}$ points, since the repulsive interaction lifts the two-fold spin degeneracy, making the number of charge degrees of freedom twice that for spin in the spin-unpolarised system. (b) Evolution of the spin and charge mass as a function of $\gamma$. The symbols are calculated numerically by repeating the calculation shown in (a) for several $\gamma$ (only every third point is shown in the figure for simplicity) and the lines are obtained from smooth interpolation. (c) Dependence of the $m_\textrm{s}/m_\textrm{c}$ ratio obtained from the pair of curves in (b).}
    \label{theory}
\end{figure*}

Having shown our technique to work successfully in 2D, we now use it in order to extract $m_0$ by probing the dispersion of the 1D wires. In this section, we will discuss the theoretical model employed with the next section detailing on the experimental measurement.

In order to get some microscopic interpretation of our system and extract $m_0$, we analyse the measured 1D dispersions by comparing them with the many-body spectra as predicted by the 1D Fermi-Hubbard model. Here,
\begin{equation}
    H=-t\sum^{L/a}_{j=1,\alpha=\uparrow,\downarrow}\big(c^\dagger_{j\alpha}c_{j+1,\alpha}+c^\dagger_{j\alpha}c_{j-1,\alpha}\big)+U\sum^{L/a}_{j=1}n_{j\uparrow}n_{j\downarrow},
\end{equation}
where $c_{j\alpha}$ are the Fermi ladder operators, $\alpha$ is the spin index $\uparrow$ or $\downarrow$, $n_{j\alpha}=c^\dagger_{j\alpha}c_{j\alpha}$ the density operator, $t$ the hopping amplitude, $U$ the interaction strength, $L$ the length of the wire, and $a$ the lattice parameter of the host crystal. The many-body spectra of this model are found from the Lieb-Wu equations \cite{LiebWu68},
\begin{align}
    k_j L-\sum^{M}_{l=1}\varphi(\lambda_m-k_j a)&=2\pi I_j,\label{lieb1}\\
    \sum^N_{j=1}\varphi(\lambda_m-k_j a)-\sum^M_{l=1}\varphi(\lambda_m/2-\lambda_l/2)&=2\pi J_m,\label{lieb2}
\end{align}
where $\varphi(x)=-2\arctan(4tx/U)$ is the two-body scattering phase. A particular set of $N$ non-equal integers $I_j$ and $M$ non-equal integers $J_m$ dictate a unique solution of this system of $N+M$ connected equations for two types of momentum states, $k_j$ for charge and $\lambda_m$ for spin degrees of freedom, giving immediately the eigenenergy of the many-body state as $E=ta^2\sum^N_{j=1}k_j^2$ and its momentum as $k=\sum^N_{j=1}k_j$. In the long-wavelength limit of our semiconductor experiment, the hopping amplitude is given by the single-particle mass $m_0$ as $t=\hbar^2/\left(2 m_0 a^2\right)$, scaling the spin and the charge spectra simultaneously by $1/m_0$.

Selecting the two sets of integers as Fermi seas [$I_j=-(N-1)/2\dots(N-1)/2, J_m=-(M-1)/2\dots(M-1)/2$] and creating linear excitations on top of them corresponds to calculating two phenomenological parameters of the low-energy field theory around the Fermi points $\pm k_\textrm{F}$ (the Tomonaga-Luttinger model) \cite{Schultz90}. Extension of these excitations away from the Fermi points provides a natural continuation of the charge/spin branches into the nonlinear region. Numerical calculation of their dispersions, shown by triangles and squares in Fig.\ \ref{theory}a, gives shapes that are close to two different parabolae (see magenta and green dashed lines), which can be described by a pair of incommensurate masses $m_\textrm{s}$ and $m_\textrm{c}$. We use these two dispersions in fitting the 1D signal and its evolution with the microscopic Hubbard parameters to extract the dependence of the two masses on the interaction strength in our experiment. Instead of $U$ we use a more natural dimensionless interaction parameter of the 1D Fermi-Hubbard model \cite{OT14}, 
\begin{equation}
\gamma=\frac{\lambda_\textrm{F}}{16a}\frac{U}{t}\frac{1}{1-\frac{1}{N}\sum_{l=1}^{N/2}\frac{\lambda_{l}^{2}\left(\infty\right)-\left(\frac{U}{4t}\right)^{2}}{\lambda_{l}^{2}\left(\infty\right)+\left(\frac{U}{4t}\right)^{2}}},\label{eq:S_gamma_discrete}
\end{equation}
where $\lambda_\textrm{F}=4L/N$ is the Fermi wavelength of the free-electron
gas and $\lambda_{l}\left(\infty\right)$
are the spin part of the solution of Eqs.\,\ref{lieb1} and \ref{lieb2} in the infinite-interaction
limit $U \rightarrow \infty$. Taking the thermodynamic limit and assuming an unpolarised Heisenberg chain (see also \cite{Orbach58} for details), we obtain $1-\sum_{l}\dots/N=1.1931$, giving:
\begin{equation}
    \gamma=0.032 \frac{\lambda_F}{a}\frac{U}{t},
\label{gamma}
\end{equation}
This serves as a more detailed counterpart of the generally used interaction parameter $r_{\textrm s}$ in this particular dimension, by including screening effects, which can be quite sizeable in our samples \cite{vianez_observing_2021}.

In order to model the dependence of the holon, $m_\textrm{c}$, and the spinon, $m_\textrm{s}$, masses on the interaction strength, we repeat the calculation of the dispersions of these two bands based on the 1D Fermi-Hubbard model (presented in Fig.\ \ref{theory}a for $\gamma=2$) for a range of $\gamma$ from 0 to a large value. Fitting  two parabolae to the numerically obtained dispersions for each calculation, we find the two masses' dependence on $\gamma$, shown in Fig.\ \ref{theory}b. At very large interaction strengths (\emph{i.e.}, large $\gamma$), the masses are very different from one another, with the ratio of $m_\textrm{s}/m_\textrm{c}$ becoming infinite for $\gamma\rightarrow\infty$, since the spinon dispersion flattens out, yielding $m_\textrm{s}\rightarrow\infty$, while the holon mass remains finite in this limit, see Fig.\ \ref{theory}c. For small $\gamma$, on the other hand, the two masses are close to one another, becoming degenerate and equal to the single-particle mass ({\emph i.e.}, $m_\textrm{s}=m_\textrm{c}=m_0$) in the free-particle limit of $\gamma=0$. Since the mass ratio is a monotonic function of $\gamma$ for all interaction strengths, we can use this dependence in order to extract $\gamma$ from the sets of experimentally measured values of $m_\textrm{c}$ and $m_\textrm{s}$.

\section*{V. Measuring $m_0$}

\begin{figure*}[t]
    \centering
	\includegraphics[width=\textwidth]{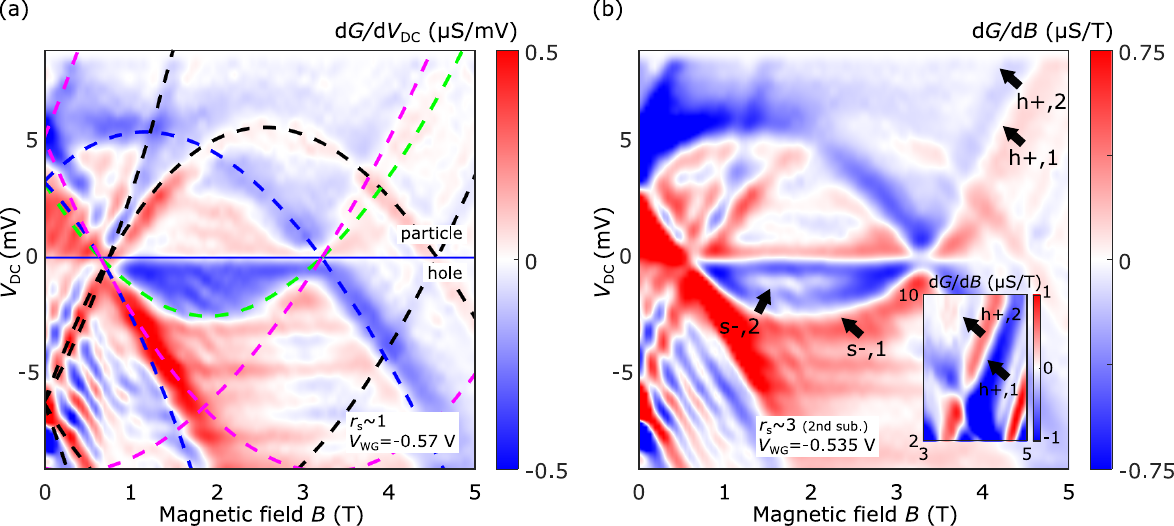}
    \caption{\textbf{Probing the dispersion of a 5\,$\mu$m 1D wire-array.}(a) Tunnelling conductance differential $\textrm{d}G/\textrm{d}V_\textrm{DC}$ \textit{vs} magnetic field $B$ and inter-layer bias $V_\textrm{DC}$ for 1D-2D tunnelling processes (dark blue region in Fig.\ \ref{fig_device}a). $V_\textrm{WG}=-0.57$\,V so that only one 1D subband is occupied. $V_\textrm{DC}>0$ corresponds to the particle sector (+), and $V_\textrm{DC}<0$ to the hole sector (-), for electrons tunnelling into and out of the wires, respectively. Black dashed lines mark the locations of the resonances resulting from the 2D-2D tunnelling processes between both wells in the `parasitic' injection region (yellow region in Fig.\ \ref{fig_device}a). This has been separately mapped and subtracted from the data shown. The blue dashed line corresponds to the dispersion of the bottom 2DEG as mapped by the 1D wires, while the green and magenta dashed lines are associated with the dispersions of the 1D system, marking the locations of the spinon (s) and holon (h) modes, respectively. (b) Same as (a) but now with $V_\textrm{WG}=-0.535$\,V so that the second 1D subband is also partially occupied. This allows us to reach significantly higher $r_\textrm{s}$ values, see text for discussion. Inset: $\textrm{d}G/\textrm{d}B$ differential of the same data showing two holon modes at high energies.}
    \label{1D_2D_map}
\end{figure*}

Fig.\ \ref{1D_2D_map}a shows a tunnelling differential map ${\rm d}G/{\rm d}V_{\rm DC}$ \textit{vs} $B$ and $V_{\rm DC}$ for a device where the wire-gate voltage $V_{\rm WG}$ is set so that only one 1D subband is occupied in the wires. The curves drawn here were, similarly to before, obtained assuming single-electron tunnelling processes between the wells, and mark the positions of resonant peaks arising from the maximal overlap of the offset spectral functions. Black dashed lines mark the location of 2D-2D resonant-tunnelling processes already separately mapped and analysed in Sec.\,III, and which have now been subtracted from the data.

From the zero-bias field intersections of the charge (magenta) and spin (green) parabolae at $B^-$ and $B^+$ (corresponding to $k=\pm k_\textrm{F}$), we extract $k_\textrm{F}=ed(B^+-B^-)/2\hbar$. This can be converted to the free-electron density $n_\textrm{1D}$ and the interaction parameter $r_\textrm{s}$, which are given by $n_\textrm{1D}=4/\lambda_\textrm{F}$, and $r_{\rm s}=1/(2a_\textrm{B}'n_\textrm{1D})$, respectively; for an equivalent analysis in 2D and 3D geometries see Appendix B. Here, $a_\textrm{B}'$ is the Bohr radius of the conduction electrons in GaAs (\textit{i.e.}, with $m_0=0.067m_\textrm{e}$ and $\epsilon\approx12$). Fitting of the whole 1D dispersion in the data reveals its modification by strong e-e interactions, including the emergence of separate collective spin and charge modes. As can be seen from the data, however, the spin parabola below the $B$-axis does not extend smoothly towards higher energies. Instead, the dispersion at positive bias extends down towards the charge line, which we interpret as indicative of the presence of two, not one, Fermi seas, for charge and for spin degrees of freedom, respectively (see our previous work \cite{vianez_observing_2021} for details). Nevertheless, both dispersion modes are essentially parabolic, meaning that they can be associated with an effective mass, $m_{\rm s}$ and $m_{\rm c}$ respectively, as is predicted by the 1D Fermi-Hubbard model and shown numerically in Fig.\ \ref{theory}. 

Our goal is to extract the electron mass $m_0$ in 1D GaAs wires as a function of density. In our previous works \cite{jompol_probing_2009,tsyplyatyev_hierarchy_2015,tsyplyatyev_nature_2016,moreno_nonlinear_2016,Jin19,vianez_observing_2021}, we generally worked at a range of $r_\textrm{s}=0.8-1.5$. Larger $r_\textrm{s}$ can nevertheless be obtained by mapping near the bottom of a subband, by depleting it to as low a density as possible. The present device design allows us to vary the number of occupied 1D subbands up to four, see Appendix B for details. Ideally then, the mapping would be done at the bottom of the first 1D subband; however, at these voltages the tunnelling signal is strongly dominated by the `parasitic' 2D injection region as the entire 1D channel is near pinch-off. In addition, the presence of localised states makes this region unsuitable for good subband resolution. Similarly, fitting to the third or fourth subband proved inadequate, partially due to the proximity to the bottom of the 2D band (where the upper 2DEG under the wire gates is not fully depleted), and also due to the increase in overall map complexity as more subbands become occupied. The most reliable data were therefore obtained by mapping at the bottom of the second subband, up to $r_\textrm{s}\sim4$, see Fig.\ \ref{1D_2D_map}b for an example of a device mapped in the two-subband regime. 

\begin{figure}[h]
    \centering
	\includegraphics[width=\columnwidth]{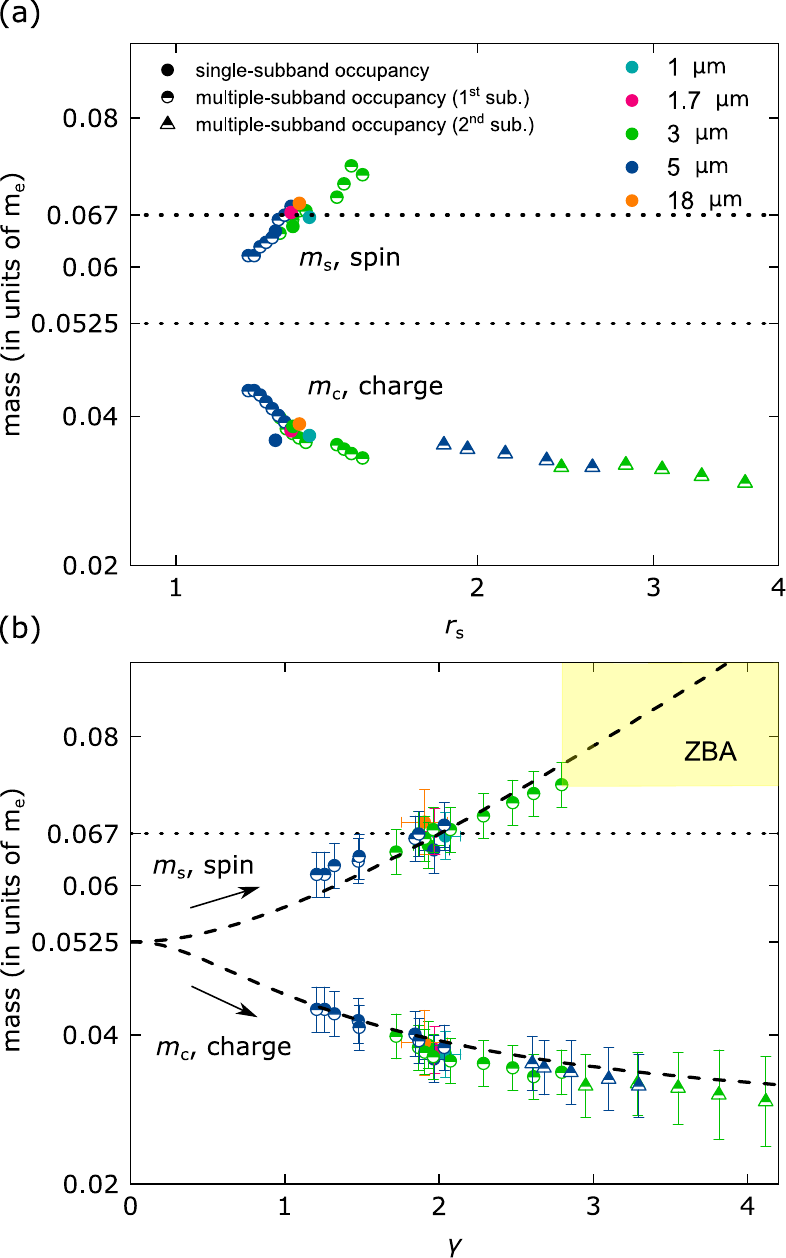}
    \caption{\textbf{Extraction of the bare electron mass $m_0$ in GaAs.} (a) Spinon ($m_\textrm{s}$) and holon ($m_\textrm{c}$) masses as a function of the interaction parameter $r_\textrm{s}$ for devices with a variety of different lengths. (b) Same data as that shown in (a) but now in terms of the interaction strength $\gamma$. Note that $\gamma$ is $\simeq r_\textrm{s}$ but also includes screening effects that are quite sizeable in our samples (see details in \cite{SM}). In a 1D geometry, $m_0$ is then given as the convergence point of these two masses in the limit as interactions are turned off (\emph{i.e.} $\gamma\rightarrow0$). Dashed curves represent a one-parameter fit for the evolution of $m_\textrm{s}$ and $m_\textrm{c}$ according to the 1D Fermi-Hubbard model. Note that the obtained value of $m_0$ is significantly below $0.067m_\textrm{e}$. The yellow shaded area marks the region in which $m_\textrm{s}$ cannot be accurately determined, due to the presence of the zero-bias anomaly (ZBA), see text for details.} 
    \label{m0_extraction}
\end{figure}

Fig.\ \ref{m0_extraction}a shows the evolution of both $m_\textrm{s}$ and $m_\textrm{c}$ on $r_\textrm{s}$. We did not observe any dependence of either mass on channel length (which was varied from 1--18\,$\mu$m). Nevertheless, in order to increase the robustness of the analysis, we focused on two samples with  longer wires (3 and 5\,$\mu $m), as they provide a larger ratio of wire to `parasitic' signal. Note that already from Fig.\ \ref{m0_extraction}a, one can already infer that the bare electron mass $m_0$ (falling somewhere in between $m_\textrm{s}$ and $m_\textrm{c}$) is significantly lower than $0.067m_\textrm{e}$.

For each measurement with a different density, we obtained the interaction strength $\gamma$ from the directly observed ratio $m_\textrm{s}/m_\textrm{c}$ using the dependence between these two quantities predicted by the 1D Fermi-Hubbard model and shown in Fig.\ \ref{theory}c. Fig.\ \ref{m0_extraction}b shows the same spin and charge mass data but now as a function of the interaction strength $\gamma$. We are able to follow the evolution of the charge mode across a large range of $\gamma$ values with good agreement with theory. We can also follow the evolution of the spin mass, as extracted from the same set of measurements, up to about $\gamma\sim3$. Above this, the spin mode is obscured by the zero-bias anomaly (ZBA), which greatly suppresses the signal within $\pm0.5$\,meV of zero bias. This is further complicated as, unlike its charge counterpart, the spin mode is only observed and tracked in the hole sector, making the extraction of $m_\textrm{s}$ more challenging. Nevertheless, the extracted values are shown to evolve systematically with $\gamma$, and they are in good agreement with our model. As the mass of each mode converges to the bare electron mass $m_0$ once interactions are turned off (\emph{i.e.}, $\gamma=0$), taking the best fit to the data as given by the 1D Fermi-Hubbard model, we obtain $m_0=(0.0525\pm0.0015)m_\textrm{e}$.

Alternatively, we can extract values of both $m_0$ and $\gamma$ from each individual measurement at a different density (\emph{i.e.}, interaction parameter $r_\textrm{s}$). In addition to extracting $\gamma$ from the observed $m_\textrm{s}/m_\textrm{c}$ ratio, we use the Hubbard spectra to fit data similar to that shown in Fig.\ \ref{1D_2D_map} by scaling the overall energy axis by $1/m_0$. As a result, we obtain an average mass of $m_0=(0.0515\pm0.0015)m_\textrm{e}$ for $r_\textrm{s}<1.6$, which shows no dependence on density (see Fig.\ \ref{m0_dimension}b, closed symbols) and is in good agreement with the previous value within experimental error. Note that uncertainty in $m_0$ mostly arises here from the error in extracting $m_\textrm{c}$ and $m_\textrm{s}$, as the observed dispersions are not perfectly sharp and have some finite broadening ($\Gamma\sim0.2-0.3$\,meV). At higher $r_\textrm{s}$ (open symbols), on the other hand, extraction of $m_\textrm{0}$ is hindered, given that $m_\textrm{s}$ cannot be accurately extracted due to the ZBA. We estimate $m_\textrm{0}$ by fitting the spinon mode up to the point where the ZBA takes over, $\gamma_{\textrm{min}}$, as well as assuming a scenario of minimal screening, $\gamma_{\textrm{max}}$, from which lower and upper bounds, respectively, on $m_0$ can be obtained given knowledge of $m_\textrm{c}$, see \cite{SM} for full details. The open symbols in Fig.\ \ref{m0_dimension}b correspond to the average values between these two limits. Therefore, although the current level of resolution of the spin mode in our experiment does not allow us to discern between different mass models, our results are compatible with a picture where $m_0$ remains constant as a function of density, as it is no longer being determined by many-body effects.

\section*{VI. Discussion}

\begin{figure}[h]
    \centering
	\includegraphics[width=\columnwidth]{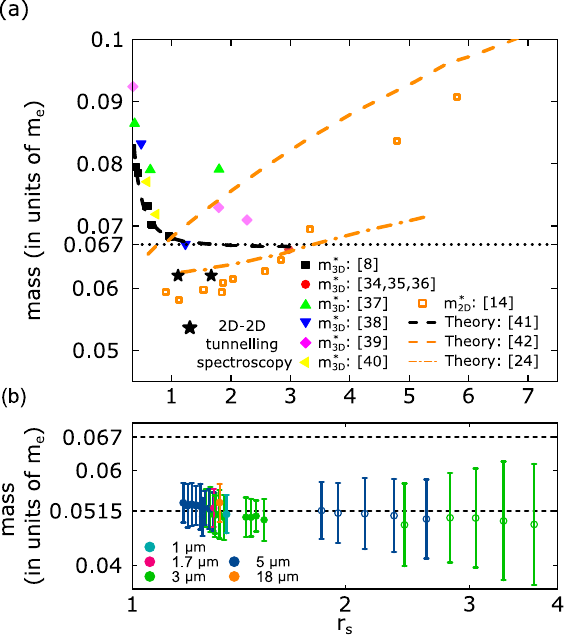}
    \caption{\textbf{Effect of dimensionality of the effective electron mass.} (a) Density dependence of the electron mass in GaAs at different dimensionalities. Three-dimensional (bulk), $m^\star_\textrm{3D}$, and two-dimensional, $m^\star_\textrm{2D}$, effective mass of electrons in GaAs as a function of interaction parameter $r_\textrm{s}$ [data taken from [8,14,24,34–42]. $\bigstar$ shows $m^\star_\textrm{2D}$ extracted from our devices using tunnelling spectroscopy, and shown in Fig.\ \ref{2D_2D_map}. (b) Bare electron mass $m_0$ extracted using our tunnelling-spectroscopy technique, for a variety of different-length devices. Closed symbols correspond to datasets where both $m_\textrm{s}$ and $m_\textrm{c}$ can be extracted, while for open symbols only $m_\textrm{c}$ is obtained.} 
    \label{m0_dimension}
\end{figure}

The value of $m_0=0.0525m_\textrm{e}$ observed in this work falls about  $22$\% below the most-commonly quoted value of the band mass, $0.067m_\textrm{e}$. A comparison with other experimental values measured at different dimensions and for various densities is presented in Fig.\ \ref{m0_dimension}a, where the data for $D>1$ are taken from \cite{raymond_electron_1979,stillman_magnetospectroscopy_1969,chamberlain_1969,hess_1976,spitzer_infrared_1959,cardona_electron_1961,piller_1966,julienne_free_1976,tan_measurements_2005,lawaetz_valence-band_1971,asgari_quasiparticle_2005,kwon_quantum_1994}. 

A systematic interpretation of this emergent picture can be given in terms of Fermi-liquid theory \cite{Landau57}, which is valid for $D>1$. Within this theory, the band mass $m_0$ is renormalised due to the many-body effect of the Coulomb interaction between electrons, producing an effective mass $m^*$.  For weak interactions (\textit{i.e.}, $r_\textrm{s}\ll 1$), the well-understood random-phase approximation \cite{Mahan_2000} gives a reduction in the effective mass $m^*/m_0=1+b_1 r_\textrm{s} \ln r_\textrm{s}+b_2 r_\textrm{s}+O( r_\textrm{s}^2)<1$, where the positive coefficients $b_i$ depend on dimensionality and details of the interaction potential, due to the screening effect, which decrease the effect of interactions. For intermediate-to-strong interactions, $r_\textrm{s}\gtrsim 1$, a larger degree of dressing in the formation of the quasi-particles competes with the screening, making the effective mass heavier ($m_*/m_0>1$), and for extremely strong interaction $r_\textrm{s}\gtrsim 20$--30 the Fermi-liquid state is expected to break down, with interacting electrons undergoing instead a type of exotic Wigner crystallisation. However, the microscopic calculation of the phenomenological parameters of the Fermi liquid for intermediate-to-strong interactions (\emph{i.e.}, $r_\textrm{s}\gtrsim 1$) is still an open problem, with effort being expended on both analytical \cite{Zhang05,Romaniello12} and numerical \cite{Giuliani08,Drummond13,Haule19} fronts. While these works converge at the qualitative level, there is as yet no firm prediction for the exact dependence of $m^*$ on $r_\textrm{s}$  beyond small $r_\textrm{s}$, and at which value of $r_\textrm{s}$ the crossover between the principal regimes occurs. 

Given this state of the theory, we can conclude from our data that for $r_\textrm{s}\simeq 1-2$ the Fermi liquid is already in the regime where the quasi-particles consist of a large-enough number of electrons to make the effective mass heavier than the single-particle mass. Analysing the dimensional dependence in Fig.\ \ref{m0_dimension}, we see that $m^*$ is heaviest for $D=3$, in which the largest coordination number makes the quasi-particles build out of the largest number of electrons geometrically. Then, $m^*$ decreases for $D=2$, as expected for a smaller coordination number, and is lightest when $D=1$, in which the phenomenon of spin-charge separation and the emergence of two separate Fermi seas fully decouples the interaction effects from the mass renormalisation, allowing the observation of the bare band mass $m_0$ directly. A further argument to support this interpretation is the strong dependence of the observed electron mass on density in $D=2,3$  but no clear variation of the mass, within the error, for the density range observed in $D=1$. Indeed, note that even if there is some dependence of $m_0$ on $r_\textrm{s}$, this can already be seen to be, within error, much weaker than that observed in 2D and 3D over a comparable range. It is also worth highlighting that the upper bound of the error bars shown for $r_\textrm{s}\gtrsim2$ was obtained assuming minimal screening, an unlikely scenario since in this region every device has two 1D subbands occupied. Therefore, the real error is most likely smaller than that shown in Fig.\ \ref{m0_dimension}b. Finally, we stress that even without applying the Hubbard model, the fact that $m_0<m^\star_\textrm{2D,3D}$ can already be seen in Fig.\ \ref{m0_extraction}a alone. For additional effects that could affect the value of $m_0$, see \cite{SM} which includes \cite{das_sarma_band_1985, pateras_mesoscopic_2018, pateras_electrode-induced_2019, Larkin97, look_electrical_1989, mancini_extended_2009-1, skinner_anomalously_2010, finkelstein_two_1993, Auslaender05, nextnano}.

\section*{VII. Conclusion}

Using the effect of spin-charge separation in 1D we have decoupled the interaction effects from the electron mass in GaAs, allowing us to measure the bare mass directly. The observed value of $m_0=0.0525m_\textrm{e}$ falls significantly below the most commonly quoted value of the band mass $0.067m_\textrm{e}$ in what is the second-most industrially important semiconductor. Our experimental findings also show that a sizeable proportion of the effective mass in 3D ($\sim22$\%) can be accounted for by interaction effects, which stresses further the need for non-perturbative methods in the microscopic development of the Fermi-liquid theory.

This result alone already provides reliable experimental data on the decoupling of the single- from the many-particle contributions to electronic parameters such as the carrier mass, which could lead to direct improvements in the modelling of materials. Simultaneously, it also opens a new opportunity for improving the operational efficiency of electronic devices, as additional control of the carrier mass can be achieved via the toolbox of many-body physics. Indeed, lower carrier mass should lead to lower resistivities, resulting in better energy efficiency, as well as faster transistors, e.g. \cite{li_very_2011}.

All data needed to evaluate the conclusions in the paper are present in the paper and/or the Supplemental Materials. The data and modelling code that support this work are also available at the University of Cambridge data repository \cite{data_availability}. 

\section{\textbf{Acknowledgments}}

We gratefully acknowledge the financial support from EPSRC (Grant No. EP/J01690X/1 and EP/J016888/1), the EPSRC International Doctoral Scholars studentship (Grant No. EP/N509620/1) and the EPSRC Doctoral Prize (P.M.T.V.), and DFG (Project No. 461313466) (O.T.).

\section{\textbf{Appendix A: Sample preparation and measurement}}

All devices measured in this work were fabricated using two double-quantum-well semiconductor heterostructures grown via molecular-beam epitaxy (MBE). These comprised two identical 18\,nm GaAs quantum wells separated by a 14\,nm Al$_{0.165}$Ga$_{0.835}$As superlattice tunnelling barrier [10 pairs of Al$_{0.33}$Ga$_{0.67}$As and GaAs monolayers]. Both wafers had 20 and 40\,nm Al$_{0.33}$Ga$_{0.67}$As spacer layers above and below the wells, respectively. These were followed in both cases by 40\,nm Si-doped layers of Al$_{0.33}$Ga$_{0.67}$As (donor concentration 1$\times10^{24}$\,m$^{-3}$). Wafer 1, however, differed from Wafer 2 by having a $100\times(2.5\,$nm$/2.5\,$nm) GaAs/AlGaAs superlattice below the $350\,$nm AlGaAs under the lower quantum well. The electron concentrations were 3 (2.2)$\times10^{15}$\,m$^{-2}$ with mobilities of 120 (165)\,m$^2$V$^{-1}$s$^{-1}$ for the top (bottom) wells for Wafer 1, and 2.85 (1.54)$\times10^{15}$\,m$^{-2}$ and 191 (55)\,m$^2$V$^{-1}$s$^{-1}$ for Wafer 2, as measured at 1.4\,K. The distance from the top of the upper well to the surface was $\sim70$\,nm, including a GaAs cap layer to prevent oxidation.

The wire gates (WGs) fabricated were 1--18\,$\mu$m-long and 0.3\,$\mu$m wide, with their separation varying between 0.15--0.17\,$\mu$m. These parameters were chosen so as to provide an energy spacing between the 1D subbands large enough that different degrees of subband filling could be probed separately. By changing the voltage applied to the WGs, one can change continuously the degree of lateral confinement, and therefore the strength of the e-e interaction. A `parasitic' gate (PG) running along the length of the array was used to modulate the density of the 2D `parasitic' injection region. All device dimensions were carefully chosen to minimise any modulation of the bottom well, which acted as a 2D spectrometer.

\begin{figure*}[t]
    \centering
	\includegraphics[width=0.95\textwidth]{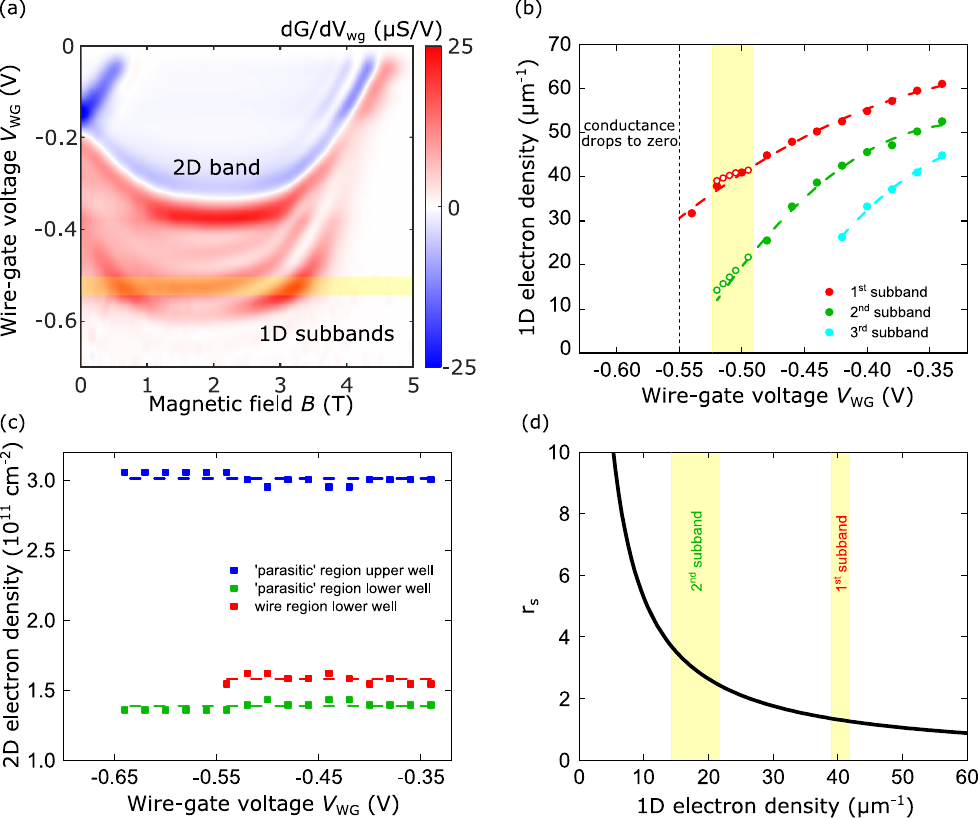}
    \caption{\textbf{Characterisation of the 1D wires and 2D spectrometer.} (a) $\textrm{d}G/\textrm{d}V_\textrm{WG}$ (where $G$ is the tunnelling conductance) as a function of both wire-gate voltage $V_\textrm{WG}$ and magnetic field $B$ for a 3\,$\mu$m device. As $V_\textrm{WG}$ becomes more negative, multiple 1D subbands start forming below the 2D band, from $V_\textrm{WG}\approx-0.3$\,V until $\approx-0.6$\,V, before the wires pinch off. The yellow shaded area marks the bottom of the second 1D subband. (b) Equilibrium 1D electron density $n_\textrm{1D}$ for each of the conducting subbands, determined from (a) (filled symbols). Open symbols correspond to the equivalent density values as extracted from the full energy-momentum maps, see text. (c) 2D electron density $n_\textrm{2D}$ of the `parasitic' injection region in both upper (blue) and lower (green) wells. The relative independence of $n_\textrm{2D}$ from $V_\textrm{WG}$, together with the proximity of the lower-well density values in the `parasitic' and wire (red) regions, allows us to use the lower well as a well-understood 2D probe (our spectrometer). (d) Interaction parameter $r_\textrm{s}$ as a function of $n_\textrm{1D}$, calculated using Eq.\ \ref{rs_parameter}}. 
    \label{sfig1}
\end{figure*}

All measurements shown in this work were carried out in a $^{3}$He cryostat at $300$\,mK. The tunnelling conductance was measured with the excitation current chosen so as to avoid sample heating. Each sample was also measured in full during a single cool-down, in order to allow for better data consistency, though different samples were independently thermally cycled, with no significant changes. In total, five different devices were measured, from two different wafers, and belonging to different fabrication batches. 

\section{\textbf{Appendix B: Interaction parameter $r_\textrm{s}$}}

The Wigner-Seitz radius $r_\textrm{s}$ is often defined as the ratio of the interaction energy to the kinetic energy, and is used as a way to estimate the interaction strength in Fermi systems independently of their dimension. It is given by
\begin{equation}
  r_\textrm{s} =
    \begin{cases}
      \frac{1}{2a_\textrm{B}n_\textrm{1D}} & \text{in 1D}\\
      \Big(\frac{1}{\pi a_\textrm{B}^2n_\textrm{2D}}\Big)^{1/2} & \text{in 2D}\\
      \Big(\frac{3}{4\pi a_\textrm{B}^3n_\textrm{3D}}\Big)^{1/3} & \text{in 3D},\\
    \end{cases}    
\label{rs_parameter}
\end{equation}
where $n_\textrm{1D}$, $n_\textrm{2D}$, and $n_\textrm{3D}$, are the respective electron densities in 1D, 2D, and 3D, and $a_\textrm{B}=4\pi\varepsilon\varepsilon_0\hbar^2/me^2$ is the Bohr radius. In GaAs, $\varepsilon\approx12$ and $m=0.067m_\textrm{e}$.

In our experiment we can vary the number of occupied 1D subbands from one to four by applying a bias $V_\textrm{WG}$ to WG, see Fig.\ \ref{sfig1}a. From here, we extract the value of the Fermi wavelength $\lambda_\textrm{F}$ in the 1D and 2D regions from their respective densities, see Fig.\ \ref{sfig1}b and \ref{sfig1}c. These values can also be obtained from maps such as that shown in Fig.\ \ref{1D_2D_map}, using the zero-bias intersection points, $B^{+\textrm{,}-}$. This gives us two independent estimates from which the value of $r_\textrm{s}$ can be extracted. Note that $n_\textrm{1D}=4/\lambda_\textrm{F}$ and $n_\textrm{2D}=2\pi/\lambda_\textrm{F}^2$.

\bibliography{citations}

\begin{thebibliography}{59}%
\makeatletter
\providecommand \@ifxundefined [1]{%
 \@ifx{#1\undefined}
}%
\providecommand \@ifnum [1]{%
 \ifnum #1\expandafter \@firstoftwo
 \else \expandafter \@secondoftwo
 \fi
}%
\providecommand \@ifx [1]{%
 \ifx #1\expandafter \@firstoftwo
 \else \expandafter \@secondoftwo
 \fi
}%
\providecommand \natexlab [1]{#1}%
\providecommand \enquote  [1]{``#1''}%
\providecommand \bibnamefont  [1]{#1}%
\providecommand \bibfnamefont [1]{#1}%
\providecommand \citenamefont [1]{#1}%
\providecommand \href@noop [0]{\@secondoftwo}%
\providecommand \href [0]{\begingroup \@sanitize@url \@href}%
\providecommand \@href[1]{\@@startlink{#1}\@@href}%
\providecommand \@@href[1]{\endgroup#1\@@endlink}%
\providecommand \@sanitize@url [0]{\catcode `\\12\catcode `\$12\catcode
  `\&12\catcode `\#12\catcode `\^12\catcode `\_12\catcode `\%12\relax}%
\providecommand \@@startlink[1]{}%
\providecommand \@@endlink[0]{}%
\providecommand \url  [0]{\begingroup\@sanitize@url \@url }%
\providecommand \@url [1]{\endgroup\@href {#1}{\urlprefix }}%
\providecommand \urlprefix  [0]{URL }%
\providecommand \Eprint [0]{\href }%
\providecommand \doibase [0]{https://doi.org/}%
\providecommand \selectlanguage [0]{\@gobble}%
\providecommand \bibinfo  [0]{\@secondoftwo}%
\providecommand \bibfield  [0]{\@secondoftwo}%
\providecommand \translation [1]{[#1]}%
\providecommand \BibitemOpen [0]{}%
\providecommand \bibitemStop [0]{}%
\providecommand \bibitemNoStop [0]{.\EOS\space}%
\providecommand \EOS [0]{\spacefactor3000\relax}%
\providecommand \BibitemShut  [1]{\csname bibitem#1\endcsname}%
\let\auto@bib@innerbib\@empty
\bibitem [{\citenamefont {Goldschmidt}(1929)}]{goldschmidt_crystal_1929}%
  \BibitemOpen
  \bibfield  {author} {\bibinfo {author} {\bibfnamefont {V.~M.}\ \bibnamefont
  {Goldschmidt}},\ }\href {https://doi.org/10.1039/TF9292500253} {\bibfield
  {journal} {\bibinfo  {journal} {Transactions of the Faraday Society}\
  }\textbf {\bibinfo {volume} {25}},\ \bibinfo {pages} {253} (\bibinfo {year}
  {1929})}\BibitemShut {NoStop}%
\bibitem [{\citenamefont {Klitzing}\ \emph {et~al.}(1980)\citenamefont
  {Klitzing}, \citenamefont {Dorda},\ and\ \citenamefont
  {Pepper}}]{klitzing_new_1980}%
  \BibitemOpen
  \bibfield  {author} {\bibinfo {author} {\bibfnamefont {K.~v.}\ \bibnamefont
  {Klitzing}}, \bibinfo {author} {\bibfnamefont {G.}~\bibnamefont {Dorda}},\
  and\ \bibinfo {author} {\bibfnamefont {M.}~\bibnamefont {Pepper}},\ }\href
  {https://doi.org/10.1103/PhysRevLett.45.494} {\bibfield  {journal} {\bibinfo
  {journal} {Phys. Rev. Lett.}\ }\textbf {\bibinfo {volume} {45}},\ \bibinfo
  {pages} {494} (\bibinfo {year} {1980})}\BibitemShut {NoStop}%
\bibitem [{\citenamefont {Tsui}\ \emph {et~al.}(1982)\citenamefont {Tsui},
  \citenamefont {Stormer},\ and\ \citenamefont
  {Gossard}}]{tsui_two-dimensional_1982}%
  \BibitemOpen
  \bibfield  {author} {\bibinfo {author} {\bibfnamefont {D.~C.}\ \bibnamefont
  {Tsui}}, \bibinfo {author} {\bibfnamefont {H.~L.}\ \bibnamefont {Stormer}},\
  and\ \bibinfo {author} {\bibfnamefont {A.~C.}\ \bibnamefont {Gossard}},\
  }\href {https://doi.org/10.1103/PhysRevLett.48.1559} {\bibfield  {journal}
  {\bibinfo  {journal} {Phys. Rev. Lett.}\ }\textbf {\bibinfo {volume} {48}},\
  \bibinfo {pages} {1559} (\bibinfo {year} {1982})}\BibitemShut {NoStop}%
\bibitem [{\citenamefont {Laughlin}(1983)}]{laughlin_anomalous_1983}%
  \BibitemOpen
  \bibfield  {author} {\bibinfo {author} {\bibfnamefont {R.~B.}\ \bibnamefont
  {Laughlin}},\ }\href {https://doi.org/10.1103/PhysRevLett.50.1395} {\bibfield
   {journal} {\bibinfo  {journal} {Phys. Rev. Lett.}\ }\textbf {\bibinfo
  {volume} {50}},\ \bibinfo {pages} {1395} (\bibinfo {year}
  {1983})}\BibitemShut {NoStop}%
\bibitem [{\citenamefont {Winkler}(2003)}]{winkler_2003}%
  \BibitemOpen
  \bibfield  {author} {\bibinfo {author} {\bibfnamefont {R.}~\bibnamefont
  {Winkler}},\ }\href {https://doi.org/10.1007/978-3-540-36616-4_11} {\emph
  {\bibinfo {title} {Spin—Orbit Coupling Effects in Two-Dimensional Electron
  and Hole Systems}}}\ (\bibinfo  {publisher} {Springer},\ \bibinfo {year}
  {2003})\BibitemShut {NoStop}%
\bibitem [{\citenamefont {Monarkha}\ and\ \citenamefont
  {Syvokon}(2012)}]{monarkha_two-dimensional_2012}%
  \BibitemOpen
  \bibfield  {author} {\bibinfo {author} {\bibfnamefont {Y.~P.}\ \bibnamefont
  {Monarkha}}\ and\ \bibinfo {author} {\bibfnamefont {V.~E.}\ \bibnamefont
  {Syvokon}},\ }\href {https://doi.org/10.1063/1.4770504} {\bibfield  {journal}
  {\bibinfo  {journal} {Low Temperature Physics}\ }\textbf {\bibinfo {volume}
  {38}},\ \bibinfo {pages} {1067} (\bibinfo {year} {2012})}\BibitemShut
  {NoStop}%
\bibitem [{\citenamefont {Manes}(2005)}]{manes_gallium_2005}%
  \BibitemOpen
  \bibfield  {author} {\bibinfo {author} {\bibfnamefont {G.~F.}\ \bibnamefont
  {Manes}},\ }\href {https://doi.org/10.1002/0471654507.eme143} {\emph
  {\bibinfo {title} {Gallium {Arsenide} {Technology} and {Applications}. In
  Encyclopedia of {RF} and {Microwave} {Engineering}}}}\ (\bibinfo  {publisher}
  {Wiley},\ \bibinfo {year} {2005})\BibitemShut {NoStop}%
\bibitem [{\citenamefont {Raymond}\ \emph {et~al.}(1979)\citenamefont
  {Raymond}, \citenamefont {Robert},\ and\ \citenamefont
  {Bernard}}]{raymond_electron_1979}%
  \BibitemOpen
  \bibfield  {author} {\bibinfo {author} {\bibfnamefont {A.}~\bibnamefont
  {Raymond}}, \bibinfo {author} {\bibfnamefont {J.~L.}\ \bibnamefont
  {Robert}},\ and\ \bibinfo {author} {\bibfnamefont {C.}~\bibnamefont
  {Bernard}},\ }\href {https://doi.org/10.1088%2F0022-3719%2F12%2F12%2F014}
  {\bibfield  {journal} {\bibinfo  {journal} {J. Phys. C: Sol. St. Phys.}\
  }\textbf {\bibinfo {volume} {12}},\ \bibinfo {pages} {2289} (\bibinfo {year}
  {1979})}\BibitemShut {NoStop}%
\bibitem [{\citenamefont {Landau}(1957)}]{Landau57}%
  \BibitemOpen
  \bibfield  {author} {\bibinfo {author} {\bibfnamefont {L.~D.}\ \bibnamefont
  {Landau}},\ }\href@noop {} {\bibfield  {journal} {\bibinfo  {journal} {Sov.
  Phys. JETP}\ }\textbf {\bibinfo {volume} {3}},\ \bibinfo {pages} {920}
  (\bibinfo {year} {1957})}\BibitemShut {NoStop}%
\bibitem [{\citenamefont {Esaki}\ and\ \citenamefont
  {Tsu}(1970)}]{esaki_superlattice_1970}%
  \BibitemOpen
  \bibfield  {author} {\bibinfo {author} {\bibfnamefont {L.}~\bibnamefont
  {Esaki}}\ and\ \bibinfo {author} {\bibfnamefont {R.}~\bibnamefont {Tsu}},\
  }\href {https://doi.org/10.1147/rd.141.0061} {\bibfield  {journal} {\bibinfo
  {journal} {IBM J. Res. Devel.}\ }\textbf {\bibinfo {volume} {14}},\ \bibinfo
  {pages} {61} (\bibinfo {year} {1970})}\BibitemShut {NoStop}%
\bibitem [{\citenamefont {Coleridge}\ \emph {et~al.}(1996)\citenamefont
  {Coleridge}, \citenamefont {Hayne}, \citenamefont {Zawadzki},\ and\
  \citenamefont {Sachrajda}}]{coleridge_effective_1996}%
  \BibitemOpen
  \bibfield  {author} {\bibinfo {author} {\bibfnamefont {P.~T.}\ \bibnamefont
  {Coleridge}}, \bibinfo {author} {\bibfnamefont {M.}~\bibnamefont {Hayne}},
  \bibinfo {author} {\bibfnamefont {P.}~\bibnamefont {Zawadzki}},\ and\
  \bibinfo {author} {\bibfnamefont {A.~S.}\ \bibnamefont {Sachrajda}},\ }\href
  {https://doi.org/10.1016/0039-6028(96)00469-4} {\bibfield  {journal}
  {\bibinfo  {journal} {Surf. Sci.}\ }\textbf {\bibinfo {volume} {361-362}},\
  \bibinfo {pages} {560} (\bibinfo {year} {1996})}\BibitemShut {NoStop}%
\bibitem [{\citenamefont {Hayne}\ \emph {et~al.}(1992)\citenamefont {Hayne},
  \citenamefont {Usher}, \citenamefont {Harris},\ and\ \citenamefont
  {Foxon}}]{hayne_exchange_1992}%
  \BibitemOpen
  \bibfield  {author} {\bibinfo {author} {\bibfnamefont {M.}~\bibnamefont
  {Hayne}}, \bibinfo {author} {\bibfnamefont {A.}~\bibnamefont {Usher}},
  \bibinfo {author} {\bibfnamefont {J.~J.}\ \bibnamefont {Harris}},\ and\
  \bibinfo {author} {\bibfnamefont {C.~T.}\ \bibnamefont {Foxon}},\ }\href
  {https://doi.org/10.1103/PhysRevB.46.9515} {\bibfield  {journal} {\bibinfo
  {journal} {Phys. Rev. B}\ }\textbf {\bibinfo {volume} {46}},\ \bibinfo
  {pages} {9515} (\bibinfo {year} {1992})}\BibitemShut {NoStop}%
\bibitem [{\citenamefont {Hatke}\ \emph {et~al.}(2013)\citenamefont {Hatke},
  \citenamefont {Zudov}, \citenamefont {Watson}, \citenamefont {Manfra},
  \citenamefont {Pfeiffer},\ and\ \citenamefont {West}}]{hatke_evidence_2013}%
  \BibitemOpen
  \bibfield  {author} {\bibinfo {author} {\bibfnamefont {A.~T.}\ \bibnamefont
  {Hatke}}, \bibinfo {author} {\bibfnamefont {M.~A.}\ \bibnamefont {Zudov}},
  \bibinfo {author} {\bibfnamefont {J.~D.}\ \bibnamefont {Watson}}, \bibinfo
  {author} {\bibfnamefont {M.~J.}\ \bibnamefont {Manfra}}, \bibinfo {author}
  {\bibfnamefont {L.~N.}\ \bibnamefont {Pfeiffer}},\ and\ \bibinfo {author}
  {\bibfnamefont {K.~W.}\ \bibnamefont {West}},\ }\href
  {https://doi.org/10.1103/PhysRevB.87.161307} {\bibfield  {journal} {\bibinfo
  {journal} {Phys. Rev. B}\ }\textbf {\bibinfo {volume} {87}},\ \bibinfo
  {pages} {161307} (\bibinfo {year} {2013})}\BibitemShut {NoStop}%
\bibitem [{\citenamefont {Tan}\ \emph {et~al.}(2005)\citenamefont {Tan},
  \citenamefont {Zhu}, \citenamefont {Stormer}, \citenamefont {Pfeiffer},
  \citenamefont {Baldwin},\ and\ \citenamefont {West}}]{tan_measurements_2005}%
  \BibitemOpen
  \bibfield  {author} {\bibinfo {author} {\bibfnamefont {Y.-W.}\ \bibnamefont
  {Tan}}, \bibinfo {author} {\bibfnamefont {J.}~\bibnamefont {Zhu}}, \bibinfo
  {author} {\bibfnamefont {H.~L.}\ \bibnamefont {Stormer}}, \bibinfo {author}
  {\bibfnamefont {L.~N.}\ \bibnamefont {Pfeiffer}}, \bibinfo {author}
  {\bibfnamefont {K.~W.}\ \bibnamefont {Baldwin}},\ and\ \bibinfo {author}
  {\bibfnamefont {K.~W.}\ \bibnamefont {West}},\ }\href
  {https://doi.org/10.1103/PhysRevLett.94.016405} {\bibfield  {journal}
  {\bibinfo  {journal} {Phys. Rev. Lett.}\ }\textbf {\bibinfo {volume} {94}},\
  \bibinfo {pages} {016405} (\bibinfo {year} {2005})}\BibitemShut {NoStop}%
\bibitem [{\citenamefont {Nilius}\ \emph {et~al.}(2002)\citenamefont {Nilius},
  \citenamefont {Wallis},\ and\ \citenamefont {Ho}}]{nilius_development_2002}%
  \BibitemOpen
  \bibfield  {author} {\bibinfo {author} {\bibfnamefont {N.}~\bibnamefont
  {Nilius}}, \bibinfo {author} {\bibfnamefont {T.~M.}\ \bibnamefont {Wallis}},\
  and\ \bibinfo {author} {\bibfnamefont {W.}~\bibnamefont {Ho}},\ }\href
  {https://doi.org/10.1126/science.1075242} {\bibfield  {journal} {\bibinfo
  {journal} {Science}\ }\textbf {\bibinfo {volume} {297}},\ \bibinfo {pages}
  {1853} (\bibinfo {year} {2002})}\BibitemShut {NoStop}%
\bibitem [{\citenamefont {Auslaender}\ \emph {et~al.}(2005)\citenamefont
  {Auslaender}, \citenamefont {Steinberg}, \citenamefont {Yacoby},
  \citenamefont {Tserkovnyak}, \citenamefont {Halperin}, \citenamefont
  {Baldwin}, \citenamefont {Pfeiffer},\ and\ \citenamefont
  {West}}]{Auslaender05}%
  \BibitemOpen
  \bibfield  {author} {\bibinfo {author} {\bibfnamefont {O.}~\bibnamefont
  {Auslaender}}, \bibinfo {author} {\bibfnamefont {H.}~\bibnamefont
  {Steinberg}}, \bibinfo {author} {\bibfnamefont {A.}~\bibnamefont {Yacoby}},
  \bibinfo {author} {\bibfnamefont {Y.}~\bibnamefont {Tserkovnyak}}, \bibinfo
  {author} {\bibfnamefont {B.~I.}\ \bibnamefont {Halperin}}, \bibinfo {author}
  {\bibfnamefont {K.~W.}\ \bibnamefont {Baldwin}}, \bibinfo {author}
  {\bibfnamefont {L.~N.}\ \bibnamefont {Pfeiffer}},\ and\ \bibinfo {author}
  {\bibfnamefont {K.~W.}\ \bibnamefont {West}},\ }\href@noop {} {\bibfield
  {journal} {\bibinfo  {journal} {Science}\ }\textbf {\bibinfo {volume}
  {308}},\ \bibinfo {pages} {88} (\bibinfo {year} {2005})}\BibitemShut
  {NoStop}%
\bibitem [{\citenamefont {Jompol}\ \emph {et~al.}(2009)\citenamefont {Jompol},
  \citenamefont {Ford}, \citenamefont {Griffiths}, \citenamefont {Farrer},
  \citenamefont {Jones}, \citenamefont {Anderson}, \citenamefont {Ritchie},
  \citenamefont {Silk},\ and\ \citenamefont {Schofield}}]{jompol_probing_2009}%
  \BibitemOpen
  \bibfield  {author} {\bibinfo {author} {\bibfnamefont {Y.}~\bibnamefont
  {Jompol}}, \bibinfo {author} {\bibfnamefont {C.~J.~B.}\ \bibnamefont {Ford}},
  \bibinfo {author} {\bibfnamefont {J.~P.}\ \bibnamefont {Griffiths}}, \bibinfo
  {author} {\bibfnamefont {I.}~\bibnamefont {Farrer}}, \bibinfo {author}
  {\bibfnamefont {G.~A.~C.}\ \bibnamefont {Jones}}, \bibinfo {author}
  {\bibfnamefont {D.}~\bibnamefont {Anderson}}, \bibinfo {author}
  {\bibfnamefont {D.~A.}\ \bibnamefont {Ritchie}}, \bibinfo {author}
  {\bibfnamefont {T.~W.}\ \bibnamefont {Silk}},\ and\ \bibinfo {author}
  {\bibfnamefont {A.~J.}\ \bibnamefont {Schofield}},\ }\href
  {https://doi.org/10.1126/science.1171769} {\bibfield  {journal} {\bibinfo
  {journal} {Science}\ }\textbf {\bibinfo {volume} {325}},\ \bibinfo {pages}
  {597} (\bibinfo {year} {2009})}\BibitemShut {NoStop}%
\bibitem [{\citenamefont {Tomonaga}(1950)}]{Tomonaga50}%
  \BibitemOpen
  \bibfield  {author} {\bibinfo {author} {\bibfnamefont {S.}~\bibnamefont
  {Tomonaga}},\ }\href@noop {} {\bibfield  {journal} {\bibinfo  {journal}
  {Prog. Theor. Phys.}\ }\textbf {\bibinfo {volume} {5}},\ \bibinfo {pages}
  {544} (\bibinfo {year} {1950})}\BibitemShut {NoStop}%
\bibitem [{\citenamefont {Luttinger}(1963)}]{Luttinger63}%
  \BibitemOpen
  \bibfield  {author} {\bibinfo {author} {\bibfnamefont {J.~M.}\ \bibnamefont
  {Luttinger}},\ }\href@noop {} {\bibfield  {journal} {\bibinfo  {journal} {J.
  Math. Phys.}\ }\textbf {\bibinfo {volume} {4}},\ \bibinfo {pages} {1154}
  (\bibinfo {year} {1963})}\BibitemShut {NoStop}%
\bibitem [{\citenamefont {Vianez}\ \emph {et~al.}(2022)\citenamefont {Vianez},
  \citenamefont {Jin}, \citenamefont {Moreno}, \citenamefont {Anirban},
  \citenamefont {Anthore}, \citenamefont {Tan}, \citenamefont {Griffiths},
  \citenamefont {Farrer}, \citenamefont {Ritchie}, \citenamefont {Schofield},
  \citenamefont {Tsyplyatyev},\ and\ \citenamefont
  {Ford}}]{vianez_observing_2021}%
  \BibitemOpen
  \bibfield  {author} {\bibinfo {author} {\bibfnamefont {P.~M.~T.}\
  \bibnamefont {Vianez}}, \bibinfo {author} {\bibfnamefont {Y.}~\bibnamefont
  {Jin}}, \bibinfo {author} {\bibfnamefont {M.}~\bibnamefont {Moreno}},
  \bibinfo {author} {\bibfnamefont {A.~S.}\ \bibnamefont {Anirban}}, \bibinfo
  {author} {\bibfnamefont {A.}~\bibnamefont {Anthore}}, \bibinfo {author}
  {\bibfnamefont {W.~K.}\ \bibnamefont {Tan}}, \bibinfo {author} {\bibfnamefont
  {J.~P.}\ \bibnamefont {Griffiths}}, \bibinfo {author} {\bibfnamefont
  {I.}~\bibnamefont {Farrer}}, \bibinfo {author} {\bibfnamefont {D.~A.}\
  \bibnamefont {Ritchie}}, \bibinfo {author} {\bibfnamefont {A.~J.}\
  \bibnamefont {Schofield}}, \bibinfo {author} {\bibfnamefont {O.}~\bibnamefont
  {Tsyplyatyev}},\ and\ \bibinfo {author} {\bibfnamefont {C.~J.~B.}\
  \bibnamefont {Ford}},\ }\href {https://doi.org/10.1126/sciadv.abm2781}
  {\bibfield  {journal} {\bibinfo  {journal} {Science Advances}\ }\textbf
  {\bibinfo {volume} {8}},\ \bibinfo {pages} {eabm2781} (\bibinfo {year}
  {2022})}\BibitemShut {NoStop}%
\bibitem [{\citenamefont {Jin}\ \emph {et~al.}(2021)\citenamefont {Jin},
  \citenamefont {Moreno}, \citenamefont {T.~Vianez}, \citenamefont {Tan},
  \citenamefont {Griffiths}, \citenamefont {Farrer}, \citenamefont {Ritchie},\
  and\ \citenamefont {Ford}}]{jin_microscopic_2021}%
  \BibitemOpen
  \bibfield  {author} {\bibinfo {author} {\bibfnamefont {Y.}~\bibnamefont
  {Jin}}, \bibinfo {author} {\bibfnamefont {M.}~\bibnamefont {Moreno}},
  \bibinfo {author} {\bibfnamefont {P.~M.}\ \bibnamefont {T.~Vianez}}, \bibinfo
  {author} {\bibfnamefont {W.~K.}\ \bibnamefont {Tan}}, \bibinfo {author}
  {\bibfnamefont {J.~P.}\ \bibnamefont {Griffiths}}, \bibinfo {author}
  {\bibfnamefont {I.}~\bibnamefont {Farrer}}, \bibinfo {author} {\bibfnamefont
  {D.~A.}\ \bibnamefont {Ritchie}},\ and\ \bibinfo {author} {\bibfnamefont
  {C.~J.~B.}\ \bibnamefont {Ford}},\ }\href@noop {} {\bibfield  {journal}
  {\bibinfo  {journal} {Appl. Phys. Lett.}\ }\textbf {\bibinfo {volume}
  {118}},\ \bibinfo {pages} {162108} (\bibinfo {year} {2021})}\BibitemShut
  {NoStop}%
\bibitem [{SM()}]{SM}%
  \BibitemOpen
  \href@noop {} {}\bibinfo {note} {See Supplementary Material at \url{https://}
  for more details.}\BibitemShut {Stop}%
\bibitem [{COM()}]{COMSOL}%
  \BibitemOpen
  \href@noop {} {}\bibinfo {note} {\textrm{https://www.comsol.com}}\BibitemShut
  {NoStop}%
\bibitem [{\citenamefont {Kwon}\ \emph {et~al.}(1994)\citenamefont {Kwon},
  \citenamefont {Ceperley},\ and\ \citenamefont {Martin}}]{kwon_quantum_1994}%
  \BibitemOpen
  \bibfield  {author} {\bibinfo {author} {\bibfnamefont {Y.}~\bibnamefont
  {Kwon}}, \bibinfo {author} {\bibfnamefont {D.~M.}\ \bibnamefont {Ceperley}},\
  and\ \bibinfo {author} {\bibfnamefont {R.~M.}\ \bibnamefont {Martin}},\
  }\href {https://doi.org/10.1103/PhysRevB.50.1684} {\bibfield  {journal}
  {\bibinfo  {journal} {Phys. Rev. B}\ }\textbf {\bibinfo {volume} {50}},\
  \bibinfo {pages} {1684} (\bibinfo {year} {1994})}\BibitemShut {NoStop}%
\bibitem [{\citenamefont {Kukushkin}\ and\ \citenamefont
  {Schmult}(2015)}]{kukushkin_fermi_2015}%
  \BibitemOpen
  \bibfield  {author} {\bibinfo {author} {\bibfnamefont {I.~V.}\ \bibnamefont
  {Kukushkin}}\ and\ \bibinfo {author} {\bibfnamefont {S.}~\bibnamefont
  {Schmult}},\ }\href {https://doi.org/10.1134/S0021364015100082} {\bibfield
  {journal} {\bibinfo  {journal} {JETP Lett.}\ }\textbf {\bibinfo {volume}
  {101}},\ \bibinfo {pages} {693} (\bibinfo {year} {2015})}\BibitemShut
  {NoStop}%
\bibitem [{\citenamefont {Lieb}\ and\ \citenamefont {Wu}(1968)}]{LiebWu68}%
  \BibitemOpen
  \bibfield  {author} {\bibinfo {author} {\bibfnamefont {E.~H.}\ \bibnamefont
  {Lieb}}\ and\ \bibinfo {author} {\bibfnamefont {F.~Y.}\ \bibnamefont {Wu}},\
  }\href@noop {} {\bibfield  {journal} {\bibinfo  {journal} {Phys. Rev. Lett.}\
  }\textbf {\bibinfo {volume} {20}},\ \bibinfo {pages} {1445} (\bibinfo {year}
  {1968})}\BibitemShut {NoStop}%
\bibitem [{\citenamefont {Schulz}(1990)}]{Schultz90}%
  \BibitemOpen
  \bibfield  {author} {\bibinfo {author} {\bibfnamefont {H.~J.}\ \bibnamefont
  {Schulz}},\ }\href@noop {} {\bibfield  {journal} {\bibinfo  {journal} {Phys.
  Rev. Lett.}\ }\textbf {\bibinfo {volume} {64}},\ \bibinfo {pages} {2831 }
  (\bibinfo {year} {1990})}\BibitemShut {NoStop}%
\bibitem [{\citenamefont {Tsyplyatyev}\ and\ \citenamefont
  {Schofield}(2014)}]{OT14}%
  \BibitemOpen
  \bibfield  {author} {\bibinfo {author} {\bibfnamefont {O.}~\bibnamefont
  {Tsyplyatyev}}\ and\ \bibinfo {author} {\bibfnamefont {A.~J.}\ \bibnamefont
  {Schofield}},\ }\href@noop {} {\bibfield  {journal} {\bibinfo  {journal}
  {Phys. Rev. B}\ }\textbf {\bibinfo {volume} {90}},\ \bibinfo {pages} {014309}
  (\bibinfo {year} {2014})}\BibitemShut {NoStop}%
\bibitem [{\citenamefont {Orbach}(1958)}]{Orbach58}%
  \BibitemOpen
  \bibfield  {author} {\bibinfo {author} {\bibfnamefont {R.}~\bibnamefont
  {Orbach}},\ }\href@noop {} {\bibfield  {journal} {\bibinfo  {journal} {Phys.
  Rev.}\ }\textbf {\bibinfo {volume} {112}},\ \bibinfo {pages} {309} (\bibinfo
  {year} {1958})}\BibitemShut {NoStop}%
\bibitem [{\citenamefont {Tsyplyatyev}\ \emph {et~al.}(2015)\citenamefont
  {Tsyplyatyev}, \citenamefont {Schofield}, \citenamefont {Jin}, \citenamefont
  {Moreno}, \citenamefont {Tan}, \citenamefont {Ford}, \citenamefont
  {Griffiths}, \citenamefont {Farrer}, \citenamefont {Jones},\ and\
  \citenamefont {Ritchie}}]{tsyplyatyev_hierarchy_2015}%
  \BibitemOpen
  \bibfield  {author} {\bibinfo {author} {\bibfnamefont {O.}~\bibnamefont
  {Tsyplyatyev}}, \bibinfo {author} {\bibfnamefont {A.~J.}\ \bibnamefont
  {Schofield}}, \bibinfo {author} {\bibfnamefont {Y.}~\bibnamefont {Jin}},
  \bibinfo {author} {\bibfnamefont {M.}~\bibnamefont {Moreno}}, \bibinfo
  {author} {\bibfnamefont {W.~K.}\ \bibnamefont {Tan}}, \bibinfo {author}
  {\bibfnamefont {C.~J.~B.}\ \bibnamefont {Ford}}, \bibinfo {author}
  {\bibfnamefont {J.~P.}\ \bibnamefont {Griffiths}}, \bibinfo {author}
  {\bibfnamefont {I.}~\bibnamefont {Farrer}}, \bibinfo {author} {\bibfnamefont
  {G.~A.~C.}\ \bibnamefont {Jones}},\ and\ \bibinfo {author} {\bibfnamefont
  {D.~A.}\ \bibnamefont {Ritchie}},\ }\href
  {https://doi.org/10.1103/PhysRevLett.114.196401} {\bibfield  {journal}
  {\bibinfo  {journal} {Phys. Rev. Lett.}\ }\textbf {\bibinfo {volume} {114}},\
  \bibinfo {pages} {196401} (\bibinfo {year} {2015})}\BibitemShut {NoStop}%
\bibitem [{\citenamefont {Tsyplyatyev}\ \emph {et~al.}(2016)\citenamefont
  {Tsyplyatyev}, \citenamefont {Schofield}, \citenamefont {Jin}, \citenamefont
  {Moreno}, \citenamefont {Tan}, \citenamefont {Anirban}, \citenamefont {Ford},
  \citenamefont {Griffiths}, \citenamefont {Farrer}, \citenamefont {Jones},\
  and\ \citenamefont {Ritchie}}]{tsyplyatyev_nature_2016}%
  \BibitemOpen
  \bibfield  {author} {\bibinfo {author} {\bibfnamefont {O.}~\bibnamefont
  {Tsyplyatyev}}, \bibinfo {author} {\bibfnamefont {A.~J.}\ \bibnamefont
  {Schofield}}, \bibinfo {author} {\bibfnamefont {Y.}~\bibnamefont {Jin}},
  \bibinfo {author} {\bibfnamefont {M.}~\bibnamefont {Moreno}}, \bibinfo
  {author} {\bibfnamefont {W.~K.}\ \bibnamefont {Tan}}, \bibinfo {author}
  {\bibfnamefont {A.~S.}\ \bibnamefont {Anirban}}, \bibinfo {author}
  {\bibfnamefont {C.~J.~B.}\ \bibnamefont {Ford}}, \bibinfo {author}
  {\bibfnamefont {J.~P.}\ \bibnamefont {Griffiths}}, \bibinfo {author}
  {\bibfnamefont {I.}~\bibnamefont {Farrer}}, \bibinfo {author} {\bibfnamefont
  {G.~A.~C.}\ \bibnamefont {Jones}},\ and\ \bibinfo {author} {\bibfnamefont
  {D.~A.}\ \bibnamefont {Ritchie}},\ }\href
  {https://doi.org/10.1103/PhysRevB.93.075147} {\bibfield  {journal} {\bibinfo
  {journal} {Phys. Rev. B}\ }\textbf {\bibinfo {volume} {93}},\ \bibinfo
  {pages} {075147} (\bibinfo {year} {2016})}\BibitemShut {NoStop}%
\bibitem [{\citenamefont {Moreno}\ \emph {et~al.}(2016)\citenamefont {Moreno},
  \citenamefont {Ford}, \citenamefont {Jin}, \citenamefont {Griffiths},
  \citenamefont {Farrer}, \citenamefont {Jones}, \citenamefont {Ritchie},
  \citenamefont {Tsyplyatyev},\ and\ \citenamefont
  {Schofield}}]{moreno_nonlinear_2016}%
  \BibitemOpen
  \bibfield  {author} {\bibinfo {author} {\bibfnamefont {M.}~\bibnamefont
  {Moreno}}, \bibinfo {author} {\bibfnamefont {C.~J.~B.}\ \bibnamefont {Ford}},
  \bibinfo {author} {\bibfnamefont {Y.}~\bibnamefont {Jin}}, \bibinfo {author}
  {\bibfnamefont {J.~P.}\ \bibnamefont {Griffiths}}, \bibinfo {author}
  {\bibfnamefont {I.}~\bibnamefont {Farrer}}, \bibinfo {author} {\bibfnamefont
  {G.~A.~C.}\ \bibnamefont {Jones}}, \bibinfo {author} {\bibfnamefont {D.~A.}\
  \bibnamefont {Ritchie}}, \bibinfo {author} {\bibfnamefont {O.}~\bibnamefont
  {Tsyplyatyev}},\ and\ \bibinfo {author} {\bibfnamefont {A.~J.}\ \bibnamefont
  {Schofield}},\ }\href {https://doi.org/10.1038/ncomms12784} {\bibfield
  {journal} {\bibinfo  {journal} {Nat. Commun.}\ }\textbf {\bibinfo {volume}
  {7}},\ \bibinfo {pages} {12784} (\bibinfo {year} {2016})}\BibitemShut
  {NoStop}%
\bibitem [{\citenamefont {Jin}\ \emph {et~al.}(2019)\citenamefont {Jin},
  \citenamefont {Tsyplyatyev}, \citenamefont {Moreno}, \citenamefont {Anthore},
  \citenamefont {Tan}, \citenamefont {Griffiths}, \citenamefont {Farrer},
  \citenamefont {Ritchie}, \citenamefont {Glazman}, \citenamefont {Schofield},\
  and\ \citenamefont {Ford}}]{Jin19}%
  \BibitemOpen
  \bibfield  {author} {\bibinfo {author} {\bibfnamefont {Y.}~\bibnamefont
  {Jin}}, \bibinfo {author} {\bibfnamefont {O.}~\bibnamefont {Tsyplyatyev}},
  \bibinfo {author} {\bibfnamefont {M.}~\bibnamefont {Moreno}}, \bibinfo
  {author} {\bibfnamefont {A.}~\bibnamefont {Anthore}}, \bibinfo {author}
  {\bibfnamefont {W.~K.}\ \bibnamefont {Tan}}, \bibinfo {author} {\bibfnamefont
  {J.~P.}\ \bibnamefont {Griffiths}}, \bibinfo {author} {\bibfnamefont
  {I.}~\bibnamefont {Farrer}}, \bibinfo {author} {\bibfnamefont {D.~A.}\
  \bibnamefont {Ritchie}}, \bibinfo {author} {\bibfnamefont {L.~I.}\
  \bibnamefont {Glazman}}, \bibinfo {author} {\bibfnamefont {A.~J.}\
  \bibnamefont {Schofield}},\ and\ \bibinfo {author} {\bibfnamefont {C.~J.~B.}\
  \bibnamefont {Ford}},\ }\href@noop {} {\bibfield  {journal} {\bibinfo
  {journal} {Nat. Commun.}\ }\textbf {\bibinfo {volume} {10}},\ \bibinfo
  {pages} {2821} (\bibinfo {year} {2019})}\BibitemShut {NoStop}%
\bibitem [{\citenamefont {Stillman}\ \emph {et~al.}(1969)\citenamefont
  {Stillman}, \citenamefont {Wolfe},\ and\ \citenamefont
  {Dimmock}}]{stillman_magnetospectroscopy_1969}%
  \BibitemOpen
  \bibfield  {author} {\bibinfo {author} {\bibfnamefont {G.~E.}\ \bibnamefont
  {Stillman}}, \bibinfo {author} {\bibfnamefont {C.~M.}\ \bibnamefont
  {Wolfe}},\ and\ \bibinfo {author} {\bibfnamefont {J.~O.}\ \bibnamefont
  {Dimmock}},\ }\href {https://doi.org/10.1016/0038-1098(69)90543-2} {\bibfield
   {journal} {\bibinfo  {journal} {Sol. State Commun.}\ }\textbf {\bibinfo
  {volume} {7}},\ \bibinfo {pages} {921} (\bibinfo {year} {1969})}\BibitemShut
  {NoStop}%
\bibitem [{\citenamefont {Chamberlain}\ \emph {et~al.}(1972)\citenamefont
  {Chamberlain}, \citenamefont {Simmonds}, \citenamefont {Stradling},\ and\
  \citenamefont {Bradley}}]{chamberlain_1969}%
  \BibitemOpen
  \bibfield  {author} {\bibinfo {author} {\bibfnamefont {J.~M.}\ \bibnamefont
  {Chamberlain}}, \bibinfo {author} {\bibfnamefont {P.~E.}\ \bibnamefont
  {Simmonds}}, \bibinfo {author} {\bibfnamefont {R.~A.}\ \bibnamefont
  {Stradling}},\ and\ \bibinfo {author} {\bibfnamefont {C.~C.}\ \bibnamefont
  {Bradley}},\ }in\ \href@noop {} {\emph {\bibinfo {booktitle} {Proc. 11th Int.
  Conf. on Physics of Semiconductors (Warsaw: Polish Scientific publishers)}}}\
  (\bibinfo {year} {1972})\ pp.\ \bibinfo {pages} {1016--1022}\BibitemShut
  {NoStop}%
\bibitem [{\citenamefont {Hess}\ \emph {et~al.}(1976)\citenamefont {Hess},
  \citenamefont {Bimberg}, \citenamefont {Lipari}, \citenamefont {Fischbach},\
  and\ \citenamefont {Altarelli}}]{hess_1976}%
  \BibitemOpen
  \bibfield  {author} {\bibinfo {author} {\bibfnamefont {K.}~\bibnamefont
  {Hess}}, \bibinfo {author} {\bibfnamefont {D.}~\bibnamefont {Bimberg}},
  \bibinfo {author} {\bibfnamefont {N.~O.}\ \bibnamefont {Lipari}}, \bibinfo
  {author} {\bibfnamefont {J.~U.}\ \bibnamefont {Fischbach}},\ and\ \bibinfo
  {author} {\bibfnamefont {M.}~\bibnamefont {Altarelli}},\ }in\ \href@noop {}
  {\emph {\bibinfo {booktitle} {Proc. 13th Int. Conf. on Physics of
  Semiconductors ed. F. G. Fumi (Rome)}}}\ (\bibinfo {year} {1976})\ pp.\
  \bibinfo {pages} {142--145}\BibitemShut {NoStop}%
\bibitem [{\citenamefont {Spitzer}\ and\ \citenamefont
  {Whelan}(1959)}]{spitzer_infrared_1959}%
  \BibitemOpen
  \bibfield  {author} {\bibinfo {author} {\bibfnamefont {W.~G.}\ \bibnamefont
  {Spitzer}}\ and\ \bibinfo {author} {\bibfnamefont {J.~M.}\ \bibnamefont
  {Whelan}},\ }\href {https://doi.org/10.1103/PhysRev.114.59} {\bibfield
  {journal} {\bibinfo  {journal} {Phys. Rev.}\ }\textbf {\bibinfo {volume}
  {114}},\ \bibinfo {pages} {59} (\bibinfo {year} {1959})}\BibitemShut
  {NoStop}%
\bibitem [{\citenamefont {Cardona}(1961)}]{cardona_electron_1961}%
  \BibitemOpen
  \bibfield  {author} {\bibinfo {author} {\bibfnamefont {M.}~\bibnamefont
  {Cardona}},\ }\href {https://doi.org/10.1103/PhysRev.121.752} {\bibfield
  {journal} {\bibinfo  {journal} {Phys. Rev.}\ }\textbf {\bibinfo {volume}
  {121}},\ \bibinfo {pages} {752} (\bibinfo {year} {1961})}\BibitemShut
  {NoStop}%
\bibitem [{\citenamefont {Piller}(1966)}]{piller_1966}%
  \BibitemOpen
  \bibfield  {author} {\bibinfo {author} {\bibfnamefont {H.}~\bibnamefont
  {Piller}},\ }in\ \href@noop {} {\emph {\bibinfo {booktitle} {Proc. 8th Int.
  Conf. on Physics of Semiconductors (Kyoto), J. Phys. Soc. Japan
  \textbf{21}}}}\ (\bibinfo {year} {1966})\ pp.\ \bibinfo {pages}
  {206--209}\BibitemShut {NoStop}%
\bibitem [{\citenamefont {Julienne}\ \emph {et~al.}(1976)\citenamefont
  {Julienne}, \citenamefont {Le~Saos}, \citenamefont {Fortini},\ and\
  \citenamefont {Bauduin}}]{julienne_free_1976}%
  \BibitemOpen
  \bibfield  {author} {\bibinfo {author} {\bibfnamefont {D.}~\bibnamefont
  {Julienne}}, \bibinfo {author} {\bibfnamefont {F.}~\bibnamefont {Le~Saos}},
  \bibinfo {author} {\bibfnamefont {A.}~\bibnamefont {Fortini}},\ and\ \bibinfo
  {author} {\bibfnamefont {P.}~\bibnamefont {Bauduin}},\ }\href
  {https://doi.org/10.1103/PhysRevB.13.2576} {\bibfield  {journal} {\bibinfo
  {journal} {Phys. Rev. B}\ }\textbf {\bibinfo {volume} {13}},\ \bibinfo
  {pages} {2576} (\bibinfo {year} {1976})}\BibitemShut {NoStop}%
\bibitem [{\citenamefont {Lawaetz}(1971)}]{lawaetz_valence-band_1971}%
  \BibitemOpen
  \bibfield  {author} {\bibinfo {author} {\bibfnamefont {P.}~\bibnamefont
  {Lawaetz}},\ }\href {https://doi.org/10.1103/PhysRevB.4.3460} {\bibfield
  {journal} {\bibinfo  {journal} {Phys. Rev. B}\ }\textbf {\bibinfo {volume}
  {4}},\ \bibinfo {pages} {3460} (\bibinfo {year} {1971})}\BibitemShut
  {NoStop}%
\bibitem [{\citenamefont {Asgari}\ \emph {et~al.}(2005)\citenamefont {Asgari},
  \citenamefont {Davoudi}, \citenamefont {Polini}, \citenamefont {Giuliani},
  \citenamefont {Tosi},\ and\ \citenamefont
  {Vignale}}]{asgari_quasiparticle_2005}%
  \BibitemOpen
  \bibfield  {author} {\bibinfo {author} {\bibfnamefont {R.}~\bibnamefont
  {Asgari}}, \bibinfo {author} {\bibfnamefont {B.}~\bibnamefont {Davoudi}},
  \bibinfo {author} {\bibfnamefont {M.}~\bibnamefont {Polini}}, \bibinfo
  {author} {\bibfnamefont {G.~F.}\ \bibnamefont {Giuliani}}, \bibinfo {author}
  {\bibfnamefont {M.~P.}\ \bibnamefont {Tosi}},\ and\ \bibinfo {author}
  {\bibfnamefont {G.}~\bibnamefont {Vignale}},\ }\href
  {https://doi.org/10.1103/PhysRevB.71.045323} {\bibfield  {journal} {\bibinfo
  {journal} {Phys. Rev. B}\ }\textbf {\bibinfo {volume} {71}},\ \bibinfo
  {pages} {045323} (\bibinfo {year} {2005})}\BibitemShut {NoStop}%
\bibitem [{\citenamefont {Mahan}(2000)}]{Mahan_2000}%
  \BibitemOpen
  \bibfield  {author} {\bibinfo {author} {\bibfnamefont {G.~D.}\ \bibnamefont
  {Mahan}},\ }\href@noop {} {\emph {\bibinfo {title} {Many-Particle Physics}}}\
  (\bibinfo  {publisher} {Springer},\ \bibinfo {year} {2000})\BibitemShut
  {NoStop}%
\bibitem [{\citenamefont {Zhang}\ and\ \citenamefont {Sarma}(2005)}]{Zhang05}%
  \BibitemOpen
  \bibfield  {author} {\bibinfo {author} {\bibfnamefont {Y.}~\bibnamefont
  {Zhang}}\ and\ \bibinfo {author} {\bibfnamefont {S.~D.}\ \bibnamefont
  {Sarma}},\ }\href@noop {} {\bibfield  {journal} {\bibinfo  {journal} {Phys.
  Rev. B}\ }\textbf {\bibinfo {volume} {71}},\ \bibinfo {pages} {045322}
  (\bibinfo {year} {2005})}\BibitemShut {NoStop}%
\bibitem [{\citenamefont {Romaniello}\ \emph {et~al.}(2012)\citenamefont
  {Romaniello}, \citenamefont {Bechstedt},\ and\ \citenamefont
  {Reining}}]{Romaniello12}%
  \BibitemOpen
  \bibfield  {author} {\bibinfo {author} {\bibfnamefont {P.}~\bibnamefont
  {Romaniello}}, \bibinfo {author} {\bibfnamefont {F.}~\bibnamefont
  {Bechstedt}},\ and\ \bibinfo {author} {\bibfnamefont {L.}~\bibnamefont
  {Reining}},\ }\href@noop {} {\bibfield  {journal} {\bibinfo  {journal} {Phys.
  Rev. B}\ }\textbf {\bibinfo {volume} {85}},\ \bibinfo {pages} {155131}
  (\bibinfo {year} {2012})}\BibitemShut {NoStop}%
\bibitem [{\citenamefont {Simion}\ and\ \citenamefont
  {Giuliani}(2008)}]{Giuliani08}%
  \BibitemOpen
  \bibfield  {author} {\bibinfo {author} {\bibfnamefont {G.~E.}\ \bibnamefont
  {Simion}}\ and\ \bibinfo {author} {\bibfnamefont {G.~F.}\ \bibnamefont
  {Giuliani}},\ }\href@noop {} {\bibfield  {journal} {\bibinfo  {journal}
  {Phys. Rev. B}\ }\textbf {\bibinfo {volume} {77}},\ \bibinfo {pages} {035131}
  (\bibinfo {year} {2008})}\BibitemShut {NoStop}%
\bibitem [{\citenamefont {Drummond}\ and\ \citenamefont
  {Needs}(2013)}]{Drummond13}%
  \BibitemOpen
  \bibfield  {author} {\bibinfo {author} {\bibfnamefont {N.~D.}\ \bibnamefont
  {Drummond}}\ and\ \bibinfo {author} {\bibfnamefont {R.~J.}\ \bibnamefont
  {Needs}},\ }\href@noop {} {\bibfield  {journal} {\bibinfo  {journal} {Phys.
  Rev. B}\ }\textbf {\bibinfo {volume} {87}},\ \bibinfo {pages} {045131}
  (\bibinfo {year} {2013})}\BibitemShut {NoStop}%
\bibitem [{\citenamefont {Chen}\ and\ \citenamefont {Haule}(2019)}]{Haule19}%
  \BibitemOpen
  \bibfield  {author} {\bibinfo {author} {\bibfnamefont {K.}~\bibnamefont
  {Chen}}\ and\ \bibinfo {author} {\bibfnamefont {K.}~\bibnamefont {Haule}},\
  }\href@noop {} {\bibfield  {journal} {\bibinfo  {journal} {Nat. Commun.}\
  }\textbf {\bibinfo {volume} {19}},\ \bibinfo {pages} {3725} (\bibinfo {year}
  {2019})}\BibitemShut {NoStop}%
\bibitem [{\citenamefont {Das~Sarma}\ and\ \citenamefont
  {Mason}(1985)}]{das_sarma_band_1985}%
  \BibitemOpen
  \bibfield  {author} {\bibinfo {author} {\bibfnamefont {S.}~\bibnamefont
  {Das~Sarma}}\ and\ \bibinfo {author} {\bibfnamefont {B.~A.}\ \bibnamefont
  {Mason}},\ }\href {https://doi.org/10.1103/PhysRevB.31.1177} {\bibfield
  {journal} {\bibinfo  {journal} {Phys. Rev. B}\ }\textbf {\bibinfo {volume}
  {31}},\ \bibinfo {pages} {1177} (\bibinfo {year} {1985})}\BibitemShut
  {NoStop}%
\bibitem [{\citenamefont {Pateras}\ \emph {et~al.}(2018)\citenamefont
  {Pateras}, \citenamefont {Park}, \citenamefont {Ahn}, \citenamefont {Tilka},
  \citenamefont {Holt}, \citenamefont {Reichl}, \citenamefont {Wegscheider},
  \citenamefont {Baart}, \citenamefont {Dehollain}, \citenamefont
  {Mukhopadhyay}, \citenamefont {Vandersypen},\ and\ \citenamefont
  {Evans}}]{pateras_mesoscopic_2018}%
  \BibitemOpen
  \bibfield  {author} {\bibinfo {author} {\bibfnamefont {A.}~\bibnamefont
  {Pateras}}, \bibinfo {author} {\bibfnamefont {J.}~\bibnamefont {Park}},
  \bibinfo {author} {\bibfnamefont {Y.}~\bibnamefont {Ahn}}, \bibinfo {author}
  {\bibfnamefont {J.~A.}\ \bibnamefont {Tilka}}, \bibinfo {author}
  {\bibfnamefont {M.~V.}\ \bibnamefont {Holt}}, \bibinfo {author}
  {\bibfnamefont {C.}~\bibnamefont {Reichl}}, \bibinfo {author} {\bibfnamefont
  {W.}~\bibnamefont {Wegscheider}}, \bibinfo {author} {\bibfnamefont {T.~A.}\
  \bibnamefont {Baart}}, \bibinfo {author} {\bibfnamefont {J.~P.}\ \bibnamefont
  {Dehollain}}, \bibinfo {author} {\bibfnamefont {U.}~\bibnamefont
  {Mukhopadhyay}}, \bibinfo {author} {\bibfnamefont {L.~M.~K.}\ \bibnamefont
  {Vandersypen}},\ and\ \bibinfo {author} {\bibfnamefont {P.~G.}\ \bibnamefont
  {Evans}},\ }\href {https://doi.org/10.1021/acs.nanolett.7b04603} {\bibfield
  {journal} {\bibinfo  {journal} {Nano Lett.}\ }\textbf {\bibinfo {volume}
  {18}},\ \bibinfo {pages} {2780} (\bibinfo {year} {2018})}\BibitemShut
  {NoStop}%
\bibitem [{\citenamefont {Pateras}\ \emph {et~al.}(2019)\citenamefont
  {Pateras}, \citenamefont {Carnis}, \citenamefont {Mukhopadhyay},
  \citenamefont {Richard}, \citenamefont {Leake}, \citenamefont {Schülli},
  \citenamefont {Reichl}, \citenamefont {Wegscheider}, \citenamefont
  {Dehollain}, \citenamefont {Vandersypen},\ and\ \citenamefont
  {Evans}}]{pateras_electrode-induced_2019}%
  \BibitemOpen
  \bibfield  {author} {\bibinfo {author} {\bibfnamefont {A.}~\bibnamefont
  {Pateras}}, \bibinfo {author} {\bibfnamefont {J.}~\bibnamefont {Carnis}},
  \bibinfo {author} {\bibfnamefont {U.}~\bibnamefont {Mukhopadhyay}}, \bibinfo
  {author} {\bibfnamefont {M.-I.}\ \bibnamefont {Richard}}, \bibinfo {author}
  {\bibfnamefont {S.~J.}\ \bibnamefont {Leake}}, \bibinfo {author}
  {\bibfnamefont {T.~U.}\ \bibnamefont {Schülli}}, \bibinfo {author}
  {\bibfnamefont {C.}~\bibnamefont {Reichl}}, \bibinfo {author} {\bibfnamefont
  {W.}~\bibnamefont {Wegscheider}}, \bibinfo {author} {\bibfnamefont {J.~P.}\
  \bibnamefont {Dehollain}}, \bibinfo {author} {\bibfnamefont {L.~M.~K.}\
  \bibnamefont {Vandersypen}},\ and\ \bibinfo {author} {\bibfnamefont {P.~G.}\
  \bibnamefont {Evans}},\ }\href {https://doi.org/10.1557/jmr.2019.61}
  {\bibfield  {journal} {\bibinfo  {journal} {J. Mat. Res.}\ }\textbf {\bibinfo
  {volume} {34}},\ \bibinfo {pages} {1291} (\bibinfo {year}
  {2019})}\BibitemShut {NoStop}%
\bibitem [{\citenamefont {Larkin}\ \emph {et~al.}(1997)\citenamefont {Larkin},
  \citenamefont {Davies}, \citenamefont {Long},\ and\ \citenamefont
  {Cusc\'o}}]{Larkin97}%
  \BibitemOpen
  \bibfield  {author} {\bibinfo {author} {\bibfnamefont {I.~A.}\ \bibnamefont
  {Larkin}}, \bibinfo {author} {\bibfnamefont {J.~H.}\ \bibnamefont {Davies}},
  \bibinfo {author} {\bibfnamefont {A.~R.}\ \bibnamefont {Long}},\ and\
  \bibinfo {author} {\bibfnamefont {R.}~\bibnamefont {Cusc\'o}},\ }\href
  {https://doi.org/10.1103/PhysRevB.56.15242} {\bibfield  {journal} {\bibinfo
  {journal} {Phys. Rev. B}\ }\textbf {\bibinfo {volume} {56}},\ \bibinfo
  {pages} {15242} (\bibinfo {year} {1997})}\BibitemShut {NoStop}%
\bibitem [{\citenamefont {Look}(1989)}]{look_electrical_1989}%
  \BibitemOpen
  \bibfield  {author} {\bibinfo {author} {\bibfnamefont {D.}~\bibnamefont
  {Look}},\ }\href {https://corescholar.libraries.wright.edu/books/32} {\emph
  {\bibinfo {title} {Electrical {Characterization} of {GaAs} {Materials} and
  {Devices}}}}\ (\bibinfo {year} {Wiley, 1989})\BibitemShut {NoStop}%
\bibitem [{\citenamefont {Mancini}\ and\ \citenamefont
  {Mancini}(2009)}]{mancini_extended_2009-1}%
  \BibitemOpen
  \bibfield  {author} {\bibinfo {author} {\bibfnamefont {F.}~\bibnamefont
  {Mancini}}\ and\ \bibinfo {author} {\bibfnamefont {F.~P.}\ \bibnamefont
  {Mancini}},\ }\href {https://doi.org/10.1140/epjb/e2008-00423-3} {\bibfield
  {journal} {\bibinfo  {journal} {The European Physical Journal B}\ }\textbf
  {\bibinfo {volume} {68}},\ \bibinfo {pages} {341} (\bibinfo {year}
  {2009})}\BibitemShut {NoStop}%
\bibitem [{\citenamefont {Skinner}\ and\ \citenamefont
  {Shklovskii}(2010)}]{skinner_anomalously_2010}%
  \BibitemOpen
  \bibfield  {author} {\bibinfo {author} {\bibfnamefont {B.}~\bibnamefont
  {Skinner}}\ and\ \bibinfo {author} {\bibfnamefont {B.~I.}\ \bibnamefont
  {Shklovskii}},\ }\href {https://doi.org/10.1103/PhysRevB.82.155111}
  {\bibfield  {journal} {\bibinfo  {journal} {Physical Review B}\ }\textbf
  {\bibinfo {volume} {82}},\ \bibinfo {pages} {155111} (\bibinfo {year}
  {2010})}\BibitemShut {NoStop}%
\bibitem [{\citenamefont {Finkel’stein}\ and\ \citenamefont
  {Larkin}(1993)}]{finkelstein_two_1993}%
  \BibitemOpen
  \bibfield  {author} {\bibinfo {author} {\bibfnamefont {A.~M.}\ \bibnamefont
  {Finkel’stein}}\ and\ \bibinfo {author} {\bibfnamefont {A.~I.}\
  \bibnamefont {Larkin}},\ }\href {https://doi.org/10.1103/PhysRevB.47.10461}
  {\bibfield  {journal} {\bibinfo  {journal} {Physical Review B}\ }\textbf
  {\bibinfo {volume} {47}},\ \bibinfo {pages} {10461} (\bibinfo {year}
  {1993})}\BibitemShut {NoStop}%
\bibitem [{nex()}]{nextnano}%
  \BibitemOpen
  \href@noop {} {}\bibinfo {note}
  {\textrm{https://www.nextnano.de}}\BibitemShut {NoStop}%
\bibitem [{\citenamefont {Li}\ \emph {et~al.}(2011)\citenamefont {Li},
  \citenamefont {Richter}, \citenamefont {Paetel}, \citenamefont {Kopp},
  \citenamefont {Mannhart},\ and\ \citenamefont {Ashoori}}]{li_very_2011}%
  \BibitemOpen
  \bibfield  {author} {\bibinfo {author} {\bibfnamefont {L.}~\bibnamefont
  {Li}}, \bibinfo {author} {\bibfnamefont {C.}~\bibnamefont {Richter}},
  \bibinfo {author} {\bibfnamefont {S.}~\bibnamefont {Paetel}}, \bibinfo
  {author} {\bibfnamefont {T.}~\bibnamefont {Kopp}}, \bibinfo {author}
  {\bibfnamefont {J.}~\bibnamefont {Mannhart}},\ and\ \bibinfo {author}
  {\bibfnamefont {R.~C.}\ \bibnamefont {Ashoori}},\ }\href
  {https://doi.org/10.1126/science.1204168} {\bibfield  {journal} {\bibinfo
  {journal} {Science}\ }\textbf {\bibinfo {volume} {332}},\ \bibinfo {pages}
  {825} (\bibinfo {year} {2011})}\BibitemShut {NoStop}%
\bibitem [{dat()}]{data_availability}%
  \BibitemOpen
  \href@noop {} {}\bibinfo {note}
  {\textrm{https://doi.org/10.17863/CAM.94403}}\BibitemShut {NoStop}%
\end{thebibliography}%


\begin{thebibliography}{11}%
\makeatletter
\providecommand \@ifxundefined [1]{%
 \@ifx{#1\undefined}
}%
\providecommand \@ifnum [1]{%
 \ifnum #1\expandafter \@firstoftwo
 \else \expandafter \@secondoftwo
 \fi
}%
\providecommand \@ifx [1]{%
 \ifx #1\expandafter \@firstoftwo
 \else \expandafter \@secondoftwo
 \fi
}%
\providecommand \natexlab [1]{#1}%
\providecommand \enquote  [1]{``#1''}%
\providecommand \bibnamefont  [1]{#1}%
\providecommand \bibfnamefont [1]{#1}%
\providecommand \citenamefont [1]{#1}%
\providecommand \href@noop [0]{\@secondoftwo}%
\providecommand \href [0]{\begingroup \@sanitize@url \@href}%
\providecommand \@href[1]{\@@startlink{#1}\@@href}%
\providecommand \@@href[1]{\endgroup#1\@@endlink}%
\providecommand \@sanitize@url [0]{\catcode `\\12\catcode `\$12\catcode
  `\&12\catcode `\#12\catcode `\^12\catcode `\_12\catcode `\%12\relax}%
\providecommand \@@startlink[1]{}%
\providecommand \@@endlink[0]{}%
\providecommand \url  [0]{\begingroup\@sanitize@url \@url }%
\providecommand \@url [1]{\endgroup\@href {#1}{\urlprefix }}%
\providecommand \urlprefix  [0]{URL }%
\providecommand \Eprint [0]{\href }%
\providecommand \doibase [0]{https://doi.org/}%
\providecommand \selectlanguage [0]{\@gobble}%
\providecommand \bibinfo  [0]{\@secondoftwo}%
\providecommand \bibfield  [0]{\@secondoftwo}%
\providecommand \translation [1]{[#1]}%
\providecommand \BibitemOpen [0]{}%
\providecommand \bibitemStop [0]{}%
\providecommand \bibitemNoStop [0]{.\EOS\space}%
\providecommand \EOS [0]{\spacefactor3000\relax}%
\providecommand \BibitemShut  [1]{\csname bibitem#1\endcsname}%
\let\auto@bib@innerbib\@empty
\bibitem [{\citenamefont {Vianez}\ \emph {et~al.}(2022)\citenamefont {Vianez},
  \citenamefont {Jin}, \citenamefont {Moreno}, \citenamefont {Anirban},
  \citenamefont {Anthore}, \citenamefont {Tan}, \citenamefont {Griffiths},
  \citenamefont {Farrer}, \citenamefont {Ritchie}, \citenamefont {Schofield},
  \citenamefont {Tsyplyatyev},\ and\ \citenamefont
  {Ford}}]{vianez_observing_2021}%
  \BibitemOpen
  \bibfield  {author} {\bibinfo {author} {\bibfnamefont {P.~M.~T.}\
  \bibnamefont {Vianez}}, \bibinfo {author} {\bibfnamefont {Y.}~\bibnamefont
  {Jin}}, \bibinfo {author} {\bibfnamefont {M.}~\bibnamefont {Moreno}},
  \bibinfo {author} {\bibfnamefont {A.~S.}\ \bibnamefont {Anirban}}, \bibinfo
  {author} {\bibfnamefont {A.}~\bibnamefont {Anthore}}, \bibinfo {author}
  {\bibfnamefont {W.~K.}\ \bibnamefont {Tan}}, \bibinfo {author} {\bibfnamefont
  {J.~P.}\ \bibnamefont {Griffiths}}, \bibinfo {author} {\bibfnamefont
  {I.}~\bibnamefont {Farrer}}, \bibinfo {author} {\bibfnamefont {D.~A.}\
  \bibnamefont {Ritchie}}, \bibinfo {author} {\bibfnamefont {A.~J.}\
  \bibnamefont {Schofield}}, \bibinfo {author} {\bibfnamefont {O.}~\bibnamefont
  {Tsyplyatyev}},\ and\ \bibinfo {author} {\bibfnamefont {C.~J.~B.}\
  \bibnamefont {Ford}},\ }\href {https://doi.org/10.1126/sciadv.abm2781}
  {\bibfield  {journal} {\bibinfo  {journal} {Science Advances}\ }\textbf
  {\bibinfo {volume} {8}},\ \bibinfo {pages} {eabm2781} (\bibinfo {year}
  {2022})}\BibitemShut {NoStop}%
\bibitem [{\citenamefont {Das~Sarma}\ and\ \citenamefont
  {Mason}(1985)}]{das_sarma_band_1985}%
  \BibitemOpen
  \bibfield  {author} {\bibinfo {author} {\bibfnamefont {S.}~\bibnamefont
  {Das~Sarma}}\ and\ \bibinfo {author} {\bibfnamefont {B.~A.}\ \bibnamefont
  {Mason}},\ }\href {https://doi.org/10.1103/PhysRevB.31.1177} {\bibfield
  {journal} {\bibinfo  {journal} {Phys. Rev. B}\ }\textbf {\bibinfo {volume}
  {31}},\ \bibinfo {pages} {1177} (\bibinfo {year} {1985})}\BibitemShut
  {NoStop}%
\bibitem [{\citenamefont {Pateras}\ \emph {et~al.}(2018)\citenamefont
  {Pateras}, \citenamefont {Park}, \citenamefont {Ahn}, \citenamefont {Tilka},
  \citenamefont {Holt}, \citenamefont {Reichl}, \citenamefont {Wegscheider},
  \citenamefont {Baart}, \citenamefont {Dehollain}, \citenamefont
  {Mukhopadhyay}, \citenamefont {Vandersypen},\ and\ \citenamefont
  {Evans}}]{pateras_mesoscopic_2018}%
  \BibitemOpen
  \bibfield  {author} {\bibinfo {author} {\bibfnamefont {A.}~\bibnamefont
  {Pateras}}, \bibinfo {author} {\bibfnamefont {J.}~\bibnamefont {Park}},
  \bibinfo {author} {\bibfnamefont {Y.}~\bibnamefont {Ahn}}, \bibinfo {author}
  {\bibfnamefont {J.~A.}\ \bibnamefont {Tilka}}, \bibinfo {author}
  {\bibfnamefont {M.~V.}\ \bibnamefont {Holt}}, \bibinfo {author}
  {\bibfnamefont {C.}~\bibnamefont {Reichl}}, \bibinfo {author} {\bibfnamefont
  {W.}~\bibnamefont {Wegscheider}}, \bibinfo {author} {\bibfnamefont {T.~A.}\
  \bibnamefont {Baart}}, \bibinfo {author} {\bibfnamefont {J.~P.}\ \bibnamefont
  {Dehollain}}, \bibinfo {author} {\bibfnamefont {U.}~\bibnamefont
  {Mukhopadhyay}}, \bibinfo {author} {\bibfnamefont {L.~M.~K.}\ \bibnamefont
  {Vandersypen}},\ and\ \bibinfo {author} {\bibfnamefont {P.~G.}\ \bibnamefont
  {Evans}},\ }\href {https://doi.org/10.1021/acs.nanolett.7b04603} {\bibfield
  {journal} {\bibinfo  {journal} {Nano Lett.}\ }\textbf {\bibinfo {volume}
  {18}},\ \bibinfo {pages} {2780} (\bibinfo {year} {2018})}\BibitemShut
  {NoStop}%
\bibitem [{\citenamefont {Pateras}\ \emph {et~al.}(2019)\citenamefont
  {Pateras}, \citenamefont {Carnis}, \citenamefont {Mukhopadhyay},
  \citenamefont {Richard}, \citenamefont {Leake}, \citenamefont {Schülli},
  \citenamefont {Reichl}, \citenamefont {Wegscheider}, \citenamefont
  {Dehollain}, \citenamefont {Vandersypen},\ and\ \citenamefont
  {Evans}}]{pateras_electrode-induced_2019}%
  \BibitemOpen
  \bibfield  {author} {\bibinfo {author} {\bibfnamefont {A.}~\bibnamefont
  {Pateras}}, \bibinfo {author} {\bibfnamefont {J.}~\bibnamefont {Carnis}},
  \bibinfo {author} {\bibfnamefont {U.}~\bibnamefont {Mukhopadhyay}}, \bibinfo
  {author} {\bibfnamefont {M.-I.}\ \bibnamefont {Richard}}, \bibinfo {author}
  {\bibfnamefont {S.~J.}\ \bibnamefont {Leake}}, \bibinfo {author}
  {\bibfnamefont {T.~U.}\ \bibnamefont {Schülli}}, \bibinfo {author}
  {\bibfnamefont {C.}~\bibnamefont {Reichl}}, \bibinfo {author} {\bibfnamefont
  {W.}~\bibnamefont {Wegscheider}}, \bibinfo {author} {\bibfnamefont {J.~P.}\
  \bibnamefont {Dehollain}}, \bibinfo {author} {\bibfnamefont {L.~M.~K.}\
  \bibnamefont {Vandersypen}},\ and\ \bibinfo {author} {\bibfnamefont {P.~G.}\
  \bibnamefont {Evans}},\ }\href {https://doi.org/10.1557/jmr.2019.61}
  {\bibfield  {journal} {\bibinfo  {journal} {J. Mat. Res.}\ }\textbf {\bibinfo
  {volume} {34}},\ \bibinfo {pages} {1291} (\bibinfo {year}
  {2019})}\BibitemShut {NoStop}%
\bibitem [{\citenamefont {Larkin}\ \emph {et~al.}(1997)\citenamefont {Larkin},
  \citenamefont {Davies}, \citenamefont {Long},\ and\ \citenamefont
  {Cusc\'o}}]{Larkin97}%
  \BibitemOpen
  \bibfield  {author} {\bibinfo {author} {\bibfnamefont {I.~A.}\ \bibnamefont
  {Larkin}}, \bibinfo {author} {\bibfnamefont {J.~H.}\ \bibnamefont {Davies}},
  \bibinfo {author} {\bibfnamefont {A.~R.}\ \bibnamefont {Long}},\ and\
  \bibinfo {author} {\bibfnamefont {R.}~\bibnamefont {Cusc\'o}},\ }\href
  {https://doi.org/10.1103/PhysRevB.56.15242} {\bibfield  {journal} {\bibinfo
  {journal} {Phys. Rev. B}\ }\textbf {\bibinfo {volume} {56}},\ \bibinfo
  {pages} {15242} (\bibinfo {year} {1997})}\BibitemShut {NoStop}%
\bibitem [{\citenamefont {Look}(1989)}]{look_electrical_1989}%
  \BibitemOpen
  \bibfield  {author} {\bibinfo {author} {\bibfnamefont {D.}~\bibnamefont
  {Look}},\ }\href {https://corescholar.libraries.wright.edu/books/32} {\emph
  {\bibinfo {title} {Electrical {Characterization} of {GaAs} {Materials} and
  {Devices}}}}\ (\bibinfo {year} {Wiley, 1989})\BibitemShut {NoStop}%
\bibitem [{\citenamefont {Mancini}\ and\ \citenamefont
  {Mancini}(2009)}]{mancini_extended_2009-1}%
  \BibitemOpen
  \bibfield  {author} {\bibinfo {author} {\bibfnamefont {F.}~\bibnamefont
  {Mancini}}\ and\ \bibinfo {author} {\bibfnamefont {F.~P.}\ \bibnamefont
  {Mancini}},\ }\href {https://doi.org/10.1140/epjb/e2008-00423-3} {\bibfield
  {journal} {\bibinfo  {journal} {The European Physical Journal B}\ }\textbf
  {\bibinfo {volume} {68}},\ \bibinfo {pages} {341} (\bibinfo {year}
  {2009})}\BibitemShut {NoStop}%
\bibitem [{\citenamefont {Skinner}\ and\ \citenamefont
  {Shklovskii}(2010)}]{skinner_anomalously_2010}%
  \BibitemOpen
  \bibfield  {author} {\bibinfo {author} {\bibfnamefont {B.}~\bibnamefont
  {Skinner}}\ and\ \bibinfo {author} {\bibfnamefont {B.~I.}\ \bibnamefont
  {Shklovskii}},\ }\href {https://doi.org/10.1103/PhysRevB.82.155111}
  {\bibfield  {journal} {\bibinfo  {journal} {Physical Review B}\ }\textbf
  {\bibinfo {volume} {82}},\ \bibinfo {pages} {155111} (\bibinfo {year}
  {2010})}\BibitemShut {NoStop}%
\bibitem [{\citenamefont {Finkel’stein}\ and\ \citenamefont
  {Larkin}(1993)}]{finkelstein_two_1993}%
  \BibitemOpen
  \bibfield  {author} {\bibinfo {author} {\bibfnamefont {A.~M.}\ \bibnamefont
  {Finkel’stein}}\ and\ \bibinfo {author} {\bibfnamefont {A.~I.}\
  \bibnamefont {Larkin}},\ }\href {https://doi.org/10.1103/PhysRevB.47.10461}
  {\bibfield  {journal} {\bibinfo  {journal} {Physical Review B}\ }\textbf
  {\bibinfo {volume} {47}},\ \bibinfo {pages} {10461} (\bibinfo {year}
  {1993})}\BibitemShut {NoStop}%
\bibitem [{\citenamefont {Auslaender}\ \emph {et~al.}(2005)\citenamefont
  {Auslaender}, \citenamefont {Steinberg}, \citenamefont {Yacoby},
  \citenamefont {Tserkovnyak}, \citenamefont {Halperin}, \citenamefont
  {Baldwin}, \citenamefont {Pfeiffer},\ and\ \citenamefont
  {West}}]{Auslaender05}%
  \BibitemOpen
  \bibfield  {author} {\bibinfo {author} {\bibfnamefont {O.}~\bibnamefont
  {Auslaender}}, \bibinfo {author} {\bibfnamefont {H.}~\bibnamefont
  {Steinberg}}, \bibinfo {author} {\bibfnamefont {A.}~\bibnamefont {Yacoby}},
  \bibinfo {author} {\bibfnamefont {Y.}~\bibnamefont {Tserkovnyak}}, \bibinfo
  {author} {\bibfnamefont {B.~I.}\ \bibnamefont {Halperin}}, \bibinfo {author}
  {\bibfnamefont {K.~W.}\ \bibnamefont {Baldwin}}, \bibinfo {author}
  {\bibfnamefont {L.~N.}\ \bibnamefont {Pfeiffer}},\ and\ \bibinfo {author}
  {\bibfnamefont {K.~W.}\ \bibnamefont {West}},\ }\href@noop {} {\bibfield
  {journal} {\bibinfo  {journal} {Science}\ }\textbf {\bibinfo {volume}
  {308}},\ \bibinfo {pages} {88} (\bibinfo {year} {2005})}\BibitemShut
  {NoStop}%
\bibitem [{nex()}]{nextnano}%
  \BibitemOpen
  \href@noop {} {}\bibinfo {note}
  {\textrm{https://www.nextnano.de}}\BibitemShut {NoStop}%
\end{thebibliography}%

\end{document}



\title{Decoupling of the many-body effects from the electron mass in GaAs by means of reduced dimensionality: Supplementary Material}

\author{P.~M.~T. Vianez}
\author{Y.~Jin}
\author{W.~K. Tan}
\author{Q.~Liu}
\author{J.~P. Griffiths}
\author{I. Farrer}
\author{D.~A. Ritchie}
\author{O. Tsyplyatyev}
\author{C.~J.~B. Ford}



\maketitle
\tableofcontents
\addtocontents{toc}{~\hfill\textbf{Page}\par}


\section{Screening effects and extraction of $m_0$ for high $r_\textrm{s}$}

At high $r_\textrm{s}$ (\emph{i.e.}, $r_\textrm{s}\geq2$), direct extraction of $m_\textrm{0}$ (see Fig.\ 5 in the main text) is hindered by the fact that $m_\textrm{s}$ cannot be accurately determined. This is because the bottom of the spinon band falls within $\pm0.5$\,meV from zero-bias, therefore being strongly suppressed by the zero-bias anomaly (ZBA). In order to establish lower and upper bounds on the value of the bare mass, we instead determine the minimum and maximum values for $\gamma$, from which an estimate of $m_0$ can be obtained, given accurate knowledge of $m_\textrm{c}$.

\begin{figure}[]
    \centering
	\includegraphics[width=\columnwidth]{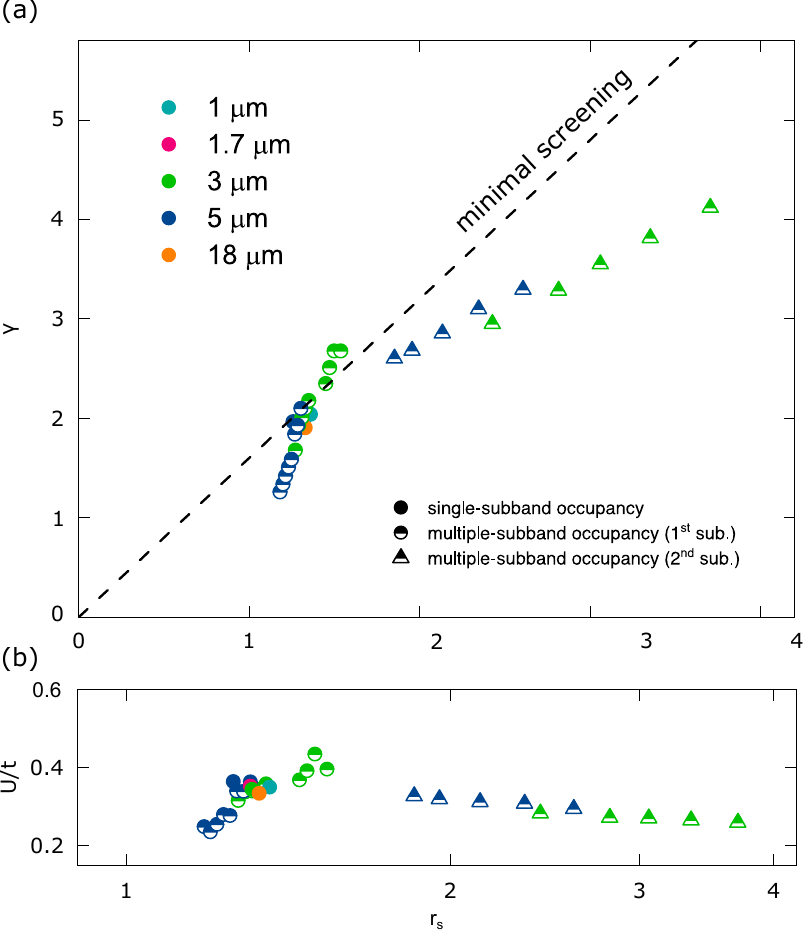}
    \caption{1D-1D inter-subband screening. (a) Interaction parameter $\gamma$ \textit{vs} $r_\textrm{s}$ for a variety of different-length devices. The dashed line corresponds to fitting using the single-subband data, and has a slope of $\sim1.6$. This marks the minimal screening boundary. Note how all other points, corresponding to two subband occupancy, fall below this line. Data points  for $r_\textrm{s} \gtrsim2$ correspond to measurements where $m_\textrm{s}$ cannot be accurately extracted due to the ZBA, see text and Fig.\ 6b in the main text for details. (b) Same data as that shown in (a) but where $U/t$ is plotted instead, see Eq.\ 9. Note that $\gamma\propto \lambda_\textrm{F} U/t$ and $r_s=\lambda_\mathrm{F}/8a_\mathrm{B}$.}
\label{sfig4}
\end{figure}

We define $\gamma_\textrm{min}$ as the minimal interaction strength allowed before the ZBA. On the other hand, $\gamma_\textrm{max}$ corresponds to the maximum interaction strength given a minimal screening regime, see Fig.\ \ref{sfig4}a. Looking exclusively at the single-subband data, we find that $\gamma$ and $r_\textrm{s}$ are approximately proportional to each other, with ratio $\approx1.6$ (dashed line), therefore marking the minimal screening boundary. All other points, corresponding to multiple-subband occupancy, systematically fall below this line. We interpret this as a manifestation of inter-subband screening, which is not captured by $r_\textrm{s}$ but is taken into account explicitly by $\gamma$ via the two-body interaction energy $U$, a result that has already been reported in \cite{vianez_observing_2021}. Therefore, $\gamma_\textrm{max}\approx 1.6 r_\textrm{s}$. Similarly, Fig.\ \ref{sfig4}b shows the same data but now in terms of $U/t$, see Eq.\,9 in the main text. Note that $U/t$ displays a stronger dependence on density at small $r_\textrm{s}$, suggesting that here screening changes more significantly than at high $r_\textrm{s}$, see also \cite{vianez_observing_2021}. From here, and given accurate knowledge of $m_\textrm{c}$, we determine $m_0^\textrm{min}$ and $m_0^\textrm{max}$, see Fig.\ 6b in the main text. The open symbols correspond to the average values between these two limits. We highlight, however, that $m_0^\textrm{max}$ is most likely an overestimate given that all data obtained at $r_\textrm{s}\geq2$ correspond to the regime with multiple subbands occupied, not just one. Nevertheless, even within error $m_0$ still falls significantly below the lowest $m_\textrm{3D}^*$ recorded in the literature. 

\section{Additional contributions to $m_0$}

There are a number of additional physical effects that could affect the extracted $m_0$ value. Below we discuss some of the most relevant, and specifically, whether or not they could lead to an enhancement or suppression of $m_0$.

\begin{itemize}
    \item \textbf{Nonparabolicity of the GaAs band:} at high densities (\textit{i.e.}, $r_\textrm{s}\lesssim1$), the nonparabolicity of the GaAs conduction band further changes the effective mass defined at the bottom of the conduction the band \cite{das_sarma_band_1985}. However, at our highest density probed this corresponds to, at most, a 5\% relative increase and therefore can largely be ignored.
    \item \textbf{Lattice strain:} lattice strain is known to lead to slight crystal distortions and, therefore, uncertainty in the position and shape of electrostatically defined structures, such as the quantum wires in the present work. In our system, this is mostly caused by the surface electrodes, as GaAs and Al$_x$Ga$_{1-x}$As are essentially lattice-matched for $x=0.3$ (mismatch $\sim4.26\times10^{-4}$). Independent work carried out using wafers and electrodes similar to our own measured a stress-induced tilt of the crystallographic planes of about 0.015$^{\circ}$, matching a stress of 57\,MPa, and therefore corresponding to a maximum in-plane strain of $\epsilon_{xx}$ of $\sim4\times10^{-5}$ \cite{pateras_mesoscopic_2018,pateras_electrode-induced_2019}. It is important to highlight, however, that these measurements were carried out at room temperature, and that it is expected that the strain is reduced by up to a factor of 2 when cooled down to cryogenic temperatures \cite{Larkin97}. Nevertheless, even when taking the maximal allowed values, they are three orders of magnitude too small to account for any significant change in the curvature of the 1D dispersion and therefore, justify the observed mass change.
    \item \textbf{Spin polarisation of the Hubbard chain:} taking the Zeeman coupling with $g=-0.44$ \cite{look_electrical_1989} for GaAs, we estimate the magnetisation in the regime of Pauli paramagnetism for a 1D Fermi gas as $g \mu_B/(2 E_F)\approx 0.03$, assuming $E_\textrm{F}=2$\,meV when in the single-subband regime. This means that at $5$\,T (\emph{i.e.} the highest field used in this experiment), the number of spin-up and spin-down electrons in the 1D chain differs by about $3$\%. Our wires can then be treated as remaining largely unpolarised under the present experimental conditions. In particular, note that this was one of the assumptions when deriving Eq.\ 9. Furthermore, modifications to the effective mass due to spin polarisation, see \cite{mancini_extended_2009-1}, are also expected to be negligible when below the half filling regime and cannot therefore explain the observed 22\% suppression in value.
   
    \item \textbf{Capacitive effects:} all our tunnelling measurements are subject to capacitive effects, not only between the two quantum wells to and from which tunnelling is occurring, but also between these and the surface gates. When extracting $m_\textrm{2D}^\star$ we discussed how this correction could be incorporated when tunnelling between two 2D systems. Similarly, when tunnelling between the 1D wires and a 2D system, we find:
    \begin{equation}k_\textrm{F}^\textrm{1D'}=k_\textrm{F}^\textrm{1D}\pm\delta k_\textrm{F}^\textrm{1D}=k_\textrm{F}^\textrm{1D}\pm\frac{\pi V_\textrm{DC}C}{2eL}.
    \end{equation}
    
    Experimentally, the capacitance values found for $C_\textrm{UW}^\textrm{2D}$ and $C_\textrm{LW}^\textrm{2D}$ are in very good agreement with the results predicted by simple COMSOL simulations. We stress, however, that in our analysis, $C_\textrm{UW}^\textrm{2D}$ and $C_\textrm{LW}^\textrm{2D}$ are free parameters and are not set by the simulation. Indeed, COMSOL ignores exchange-correlation effects which are known to modify the capacitance by up to 40\%, see \cite{skinner_anomalously_2010}, but the values are still close within the other uncertainties of the calculation.

    \begin{figure*}
    \centering
	\includegraphics[width=0.86\textwidth]{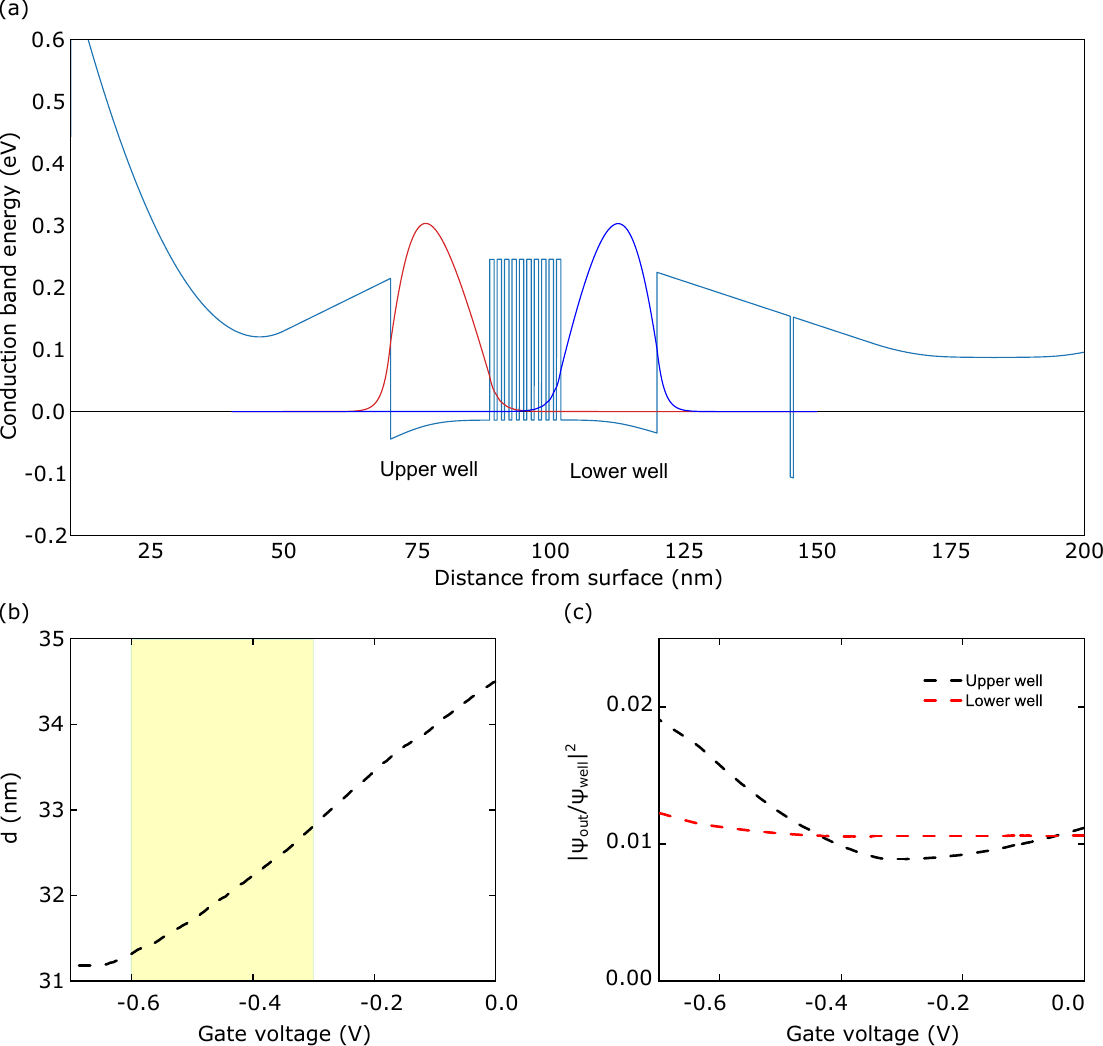}
    \caption{
    Nextnano simulation results for the semiconductor wafer material used. (a) Conduction-band structure of the GaAs/AlGaAs double-well heterostructure. Note that the top and bottom wells are roughly 80 and 110\,nm below the surface. The calculation was run assuming a $0.75$\,eV Schottky barrier at the surface.  (b) Centre-to-centre wavefunction separation $d$ as a function of surface gate voltage. Here, as seen from (a), $d\approx34$\,nm when no bias is applied. However, the wavefunctions move closer together as the gate voltage is made progressively more negative. For the typical values at which our devices are operated (shaded yellow), $d$ is expected to be around $\approx31-33$\,nm. This result is in very close agreement with our measurements where $d$ was calibrated from the 2D-2D tunnelling signal. (c) Probability of the wavefunction lying outside the well region (in the AlGaAs barriers and the superlattice tunnel barrier)---this is so small that the slightly different effective mass in the barrier makes an insignificant change to the average mass.}
\label{nextnano}
\end{figure*}

    \item \textbf{Inter-well interactions:} tunnelling between two 2DEGs, or between a wire array and a 2DEG, is a correlated process, since the separation between the electrons is comparable to the tunnelling distance (\emph{i.e.}, $\sim30$\,nm). While this is expected to modify both the tunnelling process as well as the 1D/2D spectrum, see \cite{finkelstein_two_1993}, no such modifications (gaps, extra modes, signs of hybridisation) are detected in our data. We highlight that our superlattice GaAs/AlGaAs barrier was deliberately designed so as to make the tunnelling amplitude between the two wells exponentially small, albeit still measurable with multiple wires, see Fig.\ \ref{nextnano}a for a calculation of the conduction band structure. For comparison, note that in similar tunnelling experiments between two 1D wires, see \cite{Auslaender05}, where the tunnelling barrier is less than half as wide, such modificiations were also not detected. Therefore we believe that these effects can similarly be ruled out as an explanation of the observed mass reduction.
    
    \item \textbf{Presence of a GaAs/AlGaAs barrier:} simulating the band structure of our wafer material using nextnano \cite{nextnano}, we know that the probability of the electrons in the wells penetrating into the surrounding AlGaAs barriers is about $1-2$\%, see Fig.\ \ref{nextnano}c. This would be expected to lead to an enhancement (note, not a reduction) of the extracted mass, as the effective mass in AlGaAs is slightly higher than in bulk GaAs, by at most 2\% of the difference ($0.02\times0.027=0.05\%$), which is far smaller than the other uncertainties in our result.
\end{itemize}

\bibliography{citations_SM}